\documentclass[a4paper,11pt]{article}
\usepackage[dvipdfmx]{graphicx}
\usepackage{amsmath}        
\usepackage{amssymb}        
\usepackage{bm}
\usepackage{color}
\usepackage{cite}

\makeatletter
 
 \@addtoreset{equation}{section}
\makeatother

\begin{document}

\title{Gauge Coupling Unification in Gauge-Higgs Grand Unification}

\author{Naoki Yamatsu
\footnote{Electronic address: yamatsu@het.phys.sci.osaka-u.ac.jp}
\\
{\it\small Department of Physics, Osaka University, Toyonaka, Osaka 560-0043, Japan}
}
\date{\today}

\maketitle

\begin{abstract}
We discuss renormalization group equations for gauge coupling constants
in gauge-Higgs grand unification on five-dimensional Randall-Sundrum
warped space.
We show that all the four-dimensional Standard Model gauge coupling
constants are asymptotically free and are effectively unified in $SO(11)$
gauge-Higgs grand unified theories on 5D Randall-Sundrum warped space.
\end{abstract}

\section{Introduction}

Symmetry and its breaking are essential notion in particle physics
regardless of theoretical frameworks.
The Standard Model (SM) is based on gauge symmetry 
$G_{SM}:=SU(3)_C\times SU(2)_L\times U(1)_Y$ in four-dimensional (4D)
spacetime with the spontaneous electroweak (EW) symmetry breaking 
$G_{SM}$ to $SU(3)_C\times U(1)_{em}$ via the nonvanishing vacuum
expectation value (VEV) of the SM Higgs boson.
To construct a unified theory beyond the SM, here we
use two notions; gauge-Higgs unification 
\cite{Hosotani:1983xw,Hosotani:1988bm,Davies:1987ei,Davies:1988wt,Hatanaka:1998yp} 
and grand unification
\cite{Slansky:1981yr,Yamatsu:2015gut,Georgi:1974sy,Inoue:1977qd,Fritzsch:1974nn,Ida:1980ea,Fujimoto:1981bv,Gursey:1975ki,Maekawa:2002bk,Kawamura:1999nj,Kawamura:2000ev,Kawamura:2000ir,Yamatsu:2013}.
Gauge-Higgs unification
is based on gauge symmetry in higher-dimensional spacetime. 
E.g., the $SU(3)_C\times SO(5)_W\times U(1)_X$ gauge-Higgs electroweak
(EW) unified theories on five-dimensional (5D) Randall-Sundrum (RS)
warped spacetime are discussed in
Refs.~\cite{Agashe:2004rs,Hosotani:2008tx,Funatsu:2013ni,Matsumoto:2014ila,Funatsu:2014fda,Funatsu:2014tka,Funatsu:2015xba};
the $SU(2)_L\times U(1)_Y$ EW gauge bosons and the SM Higgs
boson are unified in 5D $SO(5)_W\times U(1)_X$ bulk gauge bosons,
where the RS warped space is introduced in Ref.~\cite{Randall:1999ee}. 
Grand unification is based on grand unified (GUT) gauge symmetry.
The candidates for GUT gauge groups in 4D
GUTs are well-known. 
(See e.g., Refs.~\cite{Slansky:1981yr,Yamatsu:2015gut}.)
Also, the candidates for GUT gauge groups in 5D GUTs are shown in
Ref.~\cite{Yamatsu:2015gut}.
Gauge-Higgs grand unification 
\cite{Burdman:2002se,Haba:2002py,Haba:2004qf,Lim:2007jv,Kojima:2011ad,Frigerio:2011zg,Yamamoto:2013oja,Hosotani:2015hoa}
is base on GUT gauge symmetry in higher-dimensional spacetime.
The candidates for GUT gauge groups in 5D gauge-Higgs GUT are
shown in Ref.~\cite{Yamatsu:2015gut}. 
One of the candidates is an $SO(11)$ group. 

An $SO(11)$ gauge-Higgs grand unified theory (GHGUT) on 5D RS spacetime 
is proposed by Y.~Hosotani and the author in Ref.~\cite{Hosotani:2015hoa}.
In the $SO(11)$ GHGUT, the SM gauge bosons and the SM Higgs boson
are unified in 5D $SO(11)$ bulk gauge boson.
The SM Weyl fermions, quarks and leptons, are unified in an $SO(11)$ bulk
fermion for each generation.
Proton decay is forbidden by a fermion number conservation even if the KK
scale is much smaller than $O(10^{15})$ GeV.

In this paper, we discuss gauge coupling unification for the
4D SM gauge coupling constants of the zero modes of bulk gauge
fields in gauge-Higgs grand unification scenario, especially, $SO(11)$
GHGUTs, by using the renormalization group equations (RGEs) for the 4D
gauge coupling constants under Kaluza-Klein (KK) expansion.
We assume that the 4D description is valid until the fifth dimensional
compactification scale $1/L$ in the 5D RS warped space. 
The compactification scale $1/L$ is regarded as the real gauge coupling
unified scale $M_{GUT}$ because the $SO(11)$ GUT gauge symmetry is
broken to the $G_{PS}$ gauge symmetry by the orbifold boundary
conditions (BCs) on the Planck and TeV branes, where 
$G_{PS}:=SU(4)_C\times SU(2)_L\times SU(2)_R$ is known as 
the Pati-Salam gauge group discussed in Ref.~\cite{Pati:1974yy}.
Under the above assumption, 
we show that in several $SO(11)$ GHGUTs, the SM gauge couplings
are asymptotically free at least at one-loop level and the three SM gauge
coupling constants are almost the same values below
the GUT scale $M_{GUT}=1/L$ as long as $M_{GUT}=1/L$ is much larger than
its Kaluza-Klein (KK) mass scale 
$m_{KK}= \pi k/(e^{kL}-1)\simeq \pi k e^{-kL}$,
where $k$ is the anti-de Sitter (AdS) curvature in 5D RS warped space.

This paper is organized as follows. In Sec.~\ref{Sec:General}, we
discuss a RGE for a gauge coupling constant in 5D non-Abelian gauge
theory. In Sec.~\ref{Sec:Gauge-Higgs-grand-unification},
we discuss RGEs for the SM gauge coupling constants in the $SO(11)$
GHGUT \cite{Hosotani:2015hoa} and slightly modified ones.
We find that the three SM gauge coupling constants are
asymptotically free and they are unified in
Sec.~\ref{Sec:Asymptotic-freedom-Gauge-coupling-unification}.
Their several corrections are studied in Sec.~\ref{Sec:Corrections}.
Section~\ref{Sec:Summary-discussion} is devoted to a summary and
discussion.

\section{RGEs for 4D gauge couplings on 5D RS warped space}
\label{Sec:General}

Let us first consider a non-Abelian gauge theory on 
5D Randall-Sundrum (RS) warped spacetime.

We consider a model that contains bulk gauge and fermion fields.
Its action is given by
\begin{align}
S&=\int d^5x\sqrt{-\mbox{det}G}{\cal L}_{5D}\nonumber\\
&=\int d^5x\sqrt{-\mbox{det}G}
\left(-\frac{1}{4}\mbox{Tr}F_{MN}F^{MN}
+\overline{\Psi^{(a)}}{\cal D}(c^{(a)})\Psi^{(a)}
+{\cal L}_{g.f.}+{\cal L}_{gh}
\right),
\end{align}
where ${\cal L}_{g.f.}$ and ${\cal L}_{gh}$ stands for gauge-fixing and
ghost terms, respectively.
\begin{align}
{\cal D}_M\Psi^{(a)}(x,y)&=
\left(\partial_M-ig A_M(x,y)\right)\Psi^{(a)}(x,y),\\
A_M(x,y)&=
\frac{1}{\sqrt{2}}\sum_A A_{M}^{A}(x,y)T^{A},\\
F_{MN}(x,y)&=
\frac{i}{g}\left[{\cal D}_M,{\cal D}_N\right]
=\partial_M A_N-\partial_N A_{M}
-ig\left[A_M,A_N\right]
=\frac{1}{\sqrt{2}}\sum_A F_{MN}{}^{A}(x,y)T^{A},
\end{align}
where $M=1,2,\cdots,5$, $T^A$ are the generators of the Lie group $G$,
its superscript $A$ is the number of the generators of $G$, $\xi$ is the
gauge-fixing parameter, $g$ is the gauge coupling constant.

By using appropriate gauge-fixing and ghost terms discussed in e.g.,
Ref.~\cite{Hosotani:2009qf}, we get the KK mode expansion of the gauge
field 
\begin{align}
A_{\mu}^A(x,z)
&=\sqrt{\frac{2}{\pi R}}\sum_{n=0}^{\infty}
A_\mu^{A(n)}(x)f_n^{A}(z),\\
A_{z}^A(x,z)
&=\sqrt{\frac{2}{\pi R}}\sum_{n=0}^{\infty}
A_\mu^{A(n)}(x)h_n^{A}(z)
\end{align}
in a conformal coordinate $z:=e^{ky}$ for $|y|\leq L$, where $k$ is the
anti-de Sitter (AdS) curvature, $L$ is the size of fifth dimension,
$f_n^{A}(z)$ and $h_n^{A}(z)$ are described by using the Bessel
functions. (See, e.g., Refs.~\cite{Funatsu:2014fda,Funatsu:2014tka}.)

Here we summarize some basic results for the RGEs for 4D gauge coupling
constants. 
(See, e.g., \cite{Slansky:1981yr}.)
We only consider the RGEs at the one-loop level, but
we can find the RGEs at the two-loop level given in, e.g.,
Refs.~\cite{Machacek:1983tz,Machacek:1983fi,Machacek:1984zw}.
The RGE for the gauge coupling constant is given by 
\begin{align}
\mu\frac{dg}{d\mu}=\beta(g),
\label{Eq:RGE-gauge-coupling-4D}
\end{align}
where $\beta(g)$ is a $\beta$ function for the gauge coupling constant. 
In general, a model contains real vector, Weyl fermion, and real scalar
fields. The $\beta$ function at one-loop level is given by 
\begin{align}
\beta^{\rm 1-loop}(g)
=-\frac{g^3}{16\pi^2}\left[
\frac{11}{3}\sum_{\rm Vector}T(R_V)
-\frac{2}{3}\sum_{\rm Weyl}T(R_F)
-\frac{1}{6}\sum_{\rm Real}T(R_S)\right],
\end{align}
where Vector, Weyl, and Real stand for real vector, Weyl fermion, and
real scalar fields in terms of 4D theories, respectively.
The vector bosons are gauge bosons, so they belong to the
adjoint representation of the Lie group $G$: $T(R_V)=C_2(G)$.
$C_2(G)$ is the quadratic Casimir invariant of the adjoint
representation of $G$, and  
$T(R_i)$ is a Dynkin index of the irreducible representation $R_{i}$ of
$G$.
Note that when the Lie group $G$ is spontaneously broken into its
Lie subgroup $G'$, it is convenient to use the irreducible representations
of $G'$. (For its branching rules, see
Refs.~\cite{Yamatsu:2015gut,McKay:1981}.) 
It is convenient to use the $\beta$-function coefficient 
$b:=(16\pi^2/g^3)\beta^{\rm 1-loop}(g)$ instead 
of $\beta^{\rm 1-loop}(g)$:
\begin{align}
b=
-\frac{11}{3}\sum_{\rm Vector}T(R_V)
+\frac{2}{3}\sum_{\rm Weyl}T(R_F)
+\frac{1}{6}\sum_{\rm Real}T(R_S).
\label{Eq:beta-function-coeff-general}
\end{align}
By using $\alpha(\mu):= g^2(\mu)/4\pi$, we can rewrite the RGE in
Eq.~(\ref{Eq:RGE-gauge-coupling-4D}) as 
\begin{align}
\frac{d}{d\mbox{log}(\mu)}\alpha^{-1}(\mu)=
-\frac{b}{2\pi}.
\label{Eq:RGE-gauge-coupling-4D-alpha}
\end{align}
When $b$ is a constant, we can solve it as
\begin{align}
\alpha^{-1}(\mu)=
\alpha^{-1}(\mu_0)-\frac{b}{2\pi}\log\left(\frac{\mu}{\mu_0}\right).
\end{align}

Let us consider the RGE for 4D gauge coupling constant in 5D gauge
theories given in Eq.~(\ref{Eq:RGE-gauge-coupling-4D-alpha})
by using the $\beta$ function coefficient given in
Eq.~(\ref{Eq:beta-function-coeff-general}), where it depends on its
matter content at an energy scale $\mu$.
We take into account the contribution to the $\beta$ function
coefficient from not only zero modes but also KK modes below
their masses less than renormalization scale $\mu$, where since the
contribution to the gauge coupling constant of the zero mode from each
KK mode is almost the same as that from the zero mode, we neglect their
difference between them. 
Under the approximation, once we know mass spectra in models, we can
calculate the RGE for the gauge coupling constant at one-loop level.
In general, it is difficult to write down exact mass spectra because it
depends on orbifold boundary conditions and parameters of bulk and brane
terms. For the zeroth approximation, the mass of zero modes is $m=0$ and 
$k$-th KK modes is $m=km_{KK}$.
By using the mass spectra, the RGE of the gauge coupling constant
can be divided into two regions: 
\begin{align}
\frac{d}{d\mbox{log}(\mu)}\alpha^{-1}\simeq
\left\{
\begin{array}{ll}
-\frac{1}{2\pi}b^{0}&\ \ \mbox{for}\ \ \mu<m_{KK}\\
-\frac{1}{2\pi}
\left(b^0+k\Delta b^{KK}\right)&\ \ \mbox{for}\ \
km_{KK}\leq\mu<\left(k+1\right)m_{KK}\\
\end{array}
\right.,
\label{Eq:RGE-4D-gauge-coupling-in-5D}
\end{align}
where $b^0$ is a $\beta$-function coefficient given from its zero
modes, which can be calculated by using
Eq.~(\ref{Eq:beta-function-coeff-general});
$\Delta b^{KK}$ is an additional $\beta$-function coefficient
generated by a set of KK modes of all bulk fields, which can be also
calculated by using Eq.~(\ref{Eq:beta-function-coeff-general}).
The $\beta$-function coefficient $\Delta b^{KK}$ is
\begin{align}
\Delta b^{KK}&=-\frac{7}{2}C_2(G)+\frac{4}{3}\sum_{\rm Dirac}T(R)
\label{Eq:beta-function-coeff-KK-modes}
\end{align}
because a 5D bulk gauge field is decomposed into 4D gauge and scalar
fields and a 5D bulk fermion field is decomposed into 4D Dirac fermion
fields. 

We solve the RGE in Eq.~(\ref{Eq:RGE-4D-gauge-coupling-in-5D}).
The number of the set of KK modes for $\mu> m_{KK}$ is approximately
equal to the energy scale divided by the KK mass scale:
\begin{align}
k\simeq\frac{\mu}{m_{KK}}.
\label{Eq:Approximation}
\end{align}
We integrate the RGE in Eq.~(\ref{Eq:RGE-4D-gauge-coupling-in-5D}) 
with respect to $\mu$ from $M_Z$ to $\mu$
($M_Z<\mu<m_{KK}$):
\begin{align}
\alpha^{-1}(\mu)=&
\alpha^{-1}(M_Z)
-\frac{b^0}{2\pi}\mbox{log}\left(\frac{\mu}{M_Z}\right).
\end{align}
For $\mu> m_{KK}$, the gauge coupling constant is given by 
\begin{align}
\alpha^{-1}(\mu)\simeq&
\alpha^{-1}(m_{KK})
-\frac{b^0}{2\pi}\mbox{log}\left(\frac{\mu}{m_{KK}}\right)
-\frac{\Delta b^{KK}}{2\pi}
\left(\frac{\mu}{m_{KK}}-1\right).
\label{Eq:4D-gauge-coupling-constant-in-5D}
\end{align}

From Eq.~(\ref{Eq:4D-gauge-coupling-constant-in-5D}),
we find that for $\Delta b^{KK}>0$, 
the gauge coupling constant diverges at a certain point
\begin{align}
\alpha(\mu)\to \infty,
\label{Eq:gauge-coupling-divergence}
\end{align}
while for $\Delta b^{KK}<0$ and $\mu\gg m_{KK}$, 
the gauge coupling constant reduces rapidly:
\begin{align}
&\alpha(\mu)\simeq
\frac{-2\pi}{\Delta b^{KK}}\frac{m_{KK}}{\mu}.
\label{Eq:gauge-coupling-convergence}
\end{align}

\begin{table}[tbh]
\begin{center}
\begin{tabular}{ccccc}
\hline
Algebra&Group&Rank&$d(G)$&$C_2(G)$\\\hline
$A_n$    &$SU(n+1)$ &$n\geq 1$&$n(n+1)$ &$n+1$\\ 
$B_n$    &$SO(2n+1)$&$n\geq 3$&$n(2n+1)$&$2n-1$\\ 
$C_n$    &$USp(2n)$ &$n\geq 2$&$n(2n+1)$&$n+1$\\ 
$D_n$    &$SO(2n)$  &$n\geq 4$&$n(2n-1)$&$2(n-1)$\\ 
$E_6$    &$E_6$     &$6$      &$78$&$12$\\ 
$E_7$    &$E_7$     &$7$      &$133$&$18$\\ 
$E_8$    &$E_8$     &$8$      &$248$&$30$\\ 
$F_4$    &$F_4$     &$4$      &$52$&$9$\\ 
$G_2$    &$G_2$     &$2$      &$14$&$4$\\ 
\hline
\end{tabular}
\caption{Summary for the adjoint representation of any Lie group $G$,
where $d(G)$ and $C_2(G)$ stand for the dimension and the quadratic
Casimir invariant of the adjoint representation of $G$. 
See Refs.~\cite{Slansky:1981yr,Yamatsu:2015gut} in detail.
}
\label{Table:Summary-adjoint-representations}
\end{center}
\end{table}

From Eq.~(\ref{Eq:beta-function-coeff-KK-modes}) and the above
discussion, we also find that the gauge coupling constant of a
non-Abelian gauge field based on a simple Lie group $G$ is
asymptotically free when its matter content satisfies 
\begin{align}
\sum_{\rm Dirac}T(R)<\frac{21}{8}C_2(G)
\label{Eq:Condition-asymptotic-free}
\end{align}
because of $\Delta b^{KK}<0$.
We can check which matter content can satisfy the condition in
Eq.~(\ref{Eq:Condition-asymptotic-free}) for any classical and
exceptional Lie group by using the quadratic Casimir invariant in 
Table~\ref{Table:Summary-adjoint-representations} 
and the (second order) Dynkin index of irreducible representations of
each simple Lie group $G$ listed in
Ref.~\cite{Yamatsu:2015gut}. Especially, by using Tables in 
Appendix~A in Ref.~\cite{Yamatsu:2015gut}, it is easy to check the cases
for up to rank-15 simple Lie groups and $D_{16}=SO(32)$. Also, by using
rank-$n$ discussion, we can check it for any rank classical Lie group.

\section{Gauge-Higgs grand unification}
\label{Sec:Gauge-Higgs-grand-unification}

\begin{table}[htb]
\begin{center}
\begin{tabular}{cccc}\hline
Bulk field    &$A_M$     &$\Psi_{\bf 32}^{(a)}$&$\Psi_{\bf 11}^{(b)}$\\
\hline\hline
$SO(11)$ &${\bf 55}$&${\bf 32}$&${\bf 11}$\\
5D RS    &${\bf 5}$ &${\bf 4}$ &${\bf 4}$\\
Orbifold BC&        &$(-,-)$   &$(-,-)$\\
\hline
\end{tabular}
\hspace{2em}
\begin{tabular}{cc}\hline
Brane field        &$\phi_{\bf 16}$\\
\hline\hline
$SO(10)$          &${\bf 16}$\\
$SL(2,\mathbb{C})$&(0,0)\\
\hline
\end{tabular}
\end{center}
\caption{The matter content in the $SO(11)$ GHGUT in
Ref.~\cite{Hosotani:2015hoa}.
The left-side table shows the matter content of $SO(11)$ bulk fields.
Orbifold BC stands for the choice of signs for fermion fields.
The right-side table shows the matter content on the Planck brane.
(See Ref.~\cite{Hosotani:2015hoa} in detail.)
\label{tab:matter-content}}
\end{table}

Let us consider the RGEs for gauge coupling constants in the $SO(11)$ 
GHGUT shown in Table~\ref{tab:matter-content} and its slightly modified
ones by using the results in the previous section.
For the energy scale between $M_{Z}<\mu<m_{KK}$,
the RGEs for the SM gauge coupling constants at one-loop level are 
the same as the RGEs in the SM. 

To analyze this difference between the three SM gauge coupling
constants, we introduce the following values: 
\begin{align}
\Delta_{ij}(\mu):=&
\alpha_{i}(\mu)-\alpha_{j}(\mu),
\label{Eq:difference-gauge-coupling-1}\\
\Delta_{ij}'(\mu):=&
\alpha_{i}^{-1}(\mu)-\alpha_{j}^{-1}(\mu),
\label{Eq:difference-gauge-coupling-2}
\end{align}
where $i,j=3C,2L,1Y$ for the SM gauge coupling constants,
$\alpha_i(\mu)=g_i^2/4\pi (i=3C,2L,1Y)$, 
$\alpha_{3C}(\mu)$ is the $SU(3)_C$ gauge coupling constant,
$\alpha_{2L}(\mu)$ is the $SU(2)_L$ gauge coupling constant, and 
$\alpha_{1Y}(\mu)$ is the $U(1)_Y$ gauge coupling constant,
and we take the $SU(5)$ normalization for $U(1)_Y$.
($i,j=4C,2L,2R$ for the Pati-Salam gauge coupling constants).
From Eqs.~(\ref{Eq:difference-gauge-coupling-1}) and
(\ref{Eq:difference-gauge-coupling-2}),
we have the following relation:
\begin{align}
\Delta_{ij}(\mu)=-\Delta_{ij}'(\mu)\alpha_{i}(\mu)\alpha_{j}(\mu).
\end{align}
To discuss accuracy of unification, 
we introduce $\Xi_{ij}(\mu)$ defined by 
\begin{align}
\Xi_{ij}(\mu):=&
\frac{\Delta_{ij}(\mu)}{\alpha_{j}(\mu)}
=\frac{\alpha_i(\mu)}{\alpha_j(\mu)}-1.
\label{Eq:difference-gauge-coupling-3}
\end{align}

\begin{table}[htb]
\begin{center}
\begin{tabular}{|c|ccccccccc|}\cline{2-10}
\multicolumn{1}{c|}{}&$G_\mu$&$W_\mu$&$B_\mu$&$q$&$u^c$&$d^c$&$\ell$&$e^c$&$\phi$\\\hline\hline
$SU(3)_C$ &${\bf 8}$&${\bf 1}$&${\bf 1}$&${\bf 3}$&${\bf \overline{3}}$&${\bf \overline{3}}$&${\bf 1}$&${\bf 1}$&${\bf 2}$\\ 
$SU(2)_L$ &${\bf 1}$&${\bf 3}$&${\bf 1}$&${\bf 2}$&${\bf 1}$&${\bf 1}$&${\bf 2}$&${\bf 1}$&${\bf 2}$\\
$U(1)_Y$  &$0$&$0$&$0$&$+1/6$&$-2/3$&$+1/3$&$-1/2$&$+1$&$+1/2$\\
$SL(2,\mathbb{C})$&$\left(\frac{1}{2},\frac{1}{2}\right)$&$\left(\frac{1}{2},\frac{1}{2}\right)$&$\left(\frac{1}{2},\frac{1}{2}\right)$&$\left(\frac{1}{2},0\right)$&$\left(\frac{1}{2},0\right)$&$\left(\frac{1}{2},0\right)$&$\left(\frac{1}{2},0\right)$&$\left(\frac{1}{2},0\right)$&$(0,0)$\\[0.5em]
\hline
\end{tabular}
\end{center}
\caption{The matter content in the SM or the zero mode matter content in
the $SO(11)$ GHGUTs.
\label{tab:SM-matter-content}}
\end{table}

We check $\beta$ function coefficients of the three SM gauge
coupling constants by using the RGE in
Eq.~(\ref{Eq:beta-function-coeff-general}).
The SM matter content or the zero mode matter content in
the $SO(11)$ GHGUTs is given in Table~\ref{tab:SM-matter-content}.
By using the formula in
Eq.~(\ref{Eq:beta-function-coeff-general}) and the (second order) Dynkin
indices listed in
Refs.~\cite{McKay:1981,Slansky:1981yr,Yamatsu:2015gut}, we obtain 
the following well-known SM $\beta$-function coefficients:
\begin{align}
b_i=-\frac{11}{3}C_2(G_i)
+\frac{2}{3}\sum_{\rm Quarks\&Leptons}T(R_i)
+\frac{1}{3}\sum_{\rm Higgs}T(R_i)
=\left(
\begin{array}{c}
-7\\
-19/6\\
+41/10\\
\end{array}
\right),
\end{align}
where $i=3C,2L,1Y$ stand for $SU(3)_C$, $SU(2)_L$, $U(1)_Y$, respectively,
and we took the $SU(5)$ normalization for $U(1)_Y$.

\begin{figure}[tbh]
\begin{center}
\includegraphics[bb=0 0 322 247,height=3.5cm]{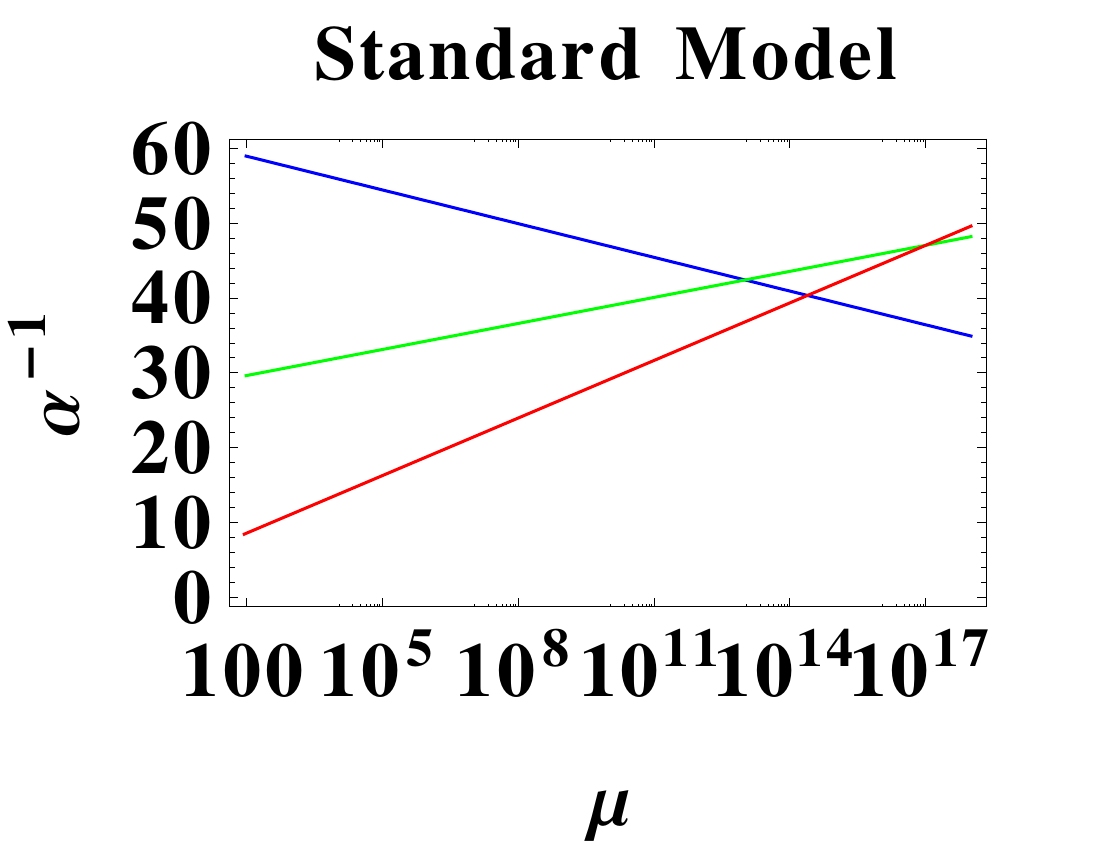}
\includegraphics[bb=0 0 335 244,height=3.5cm]{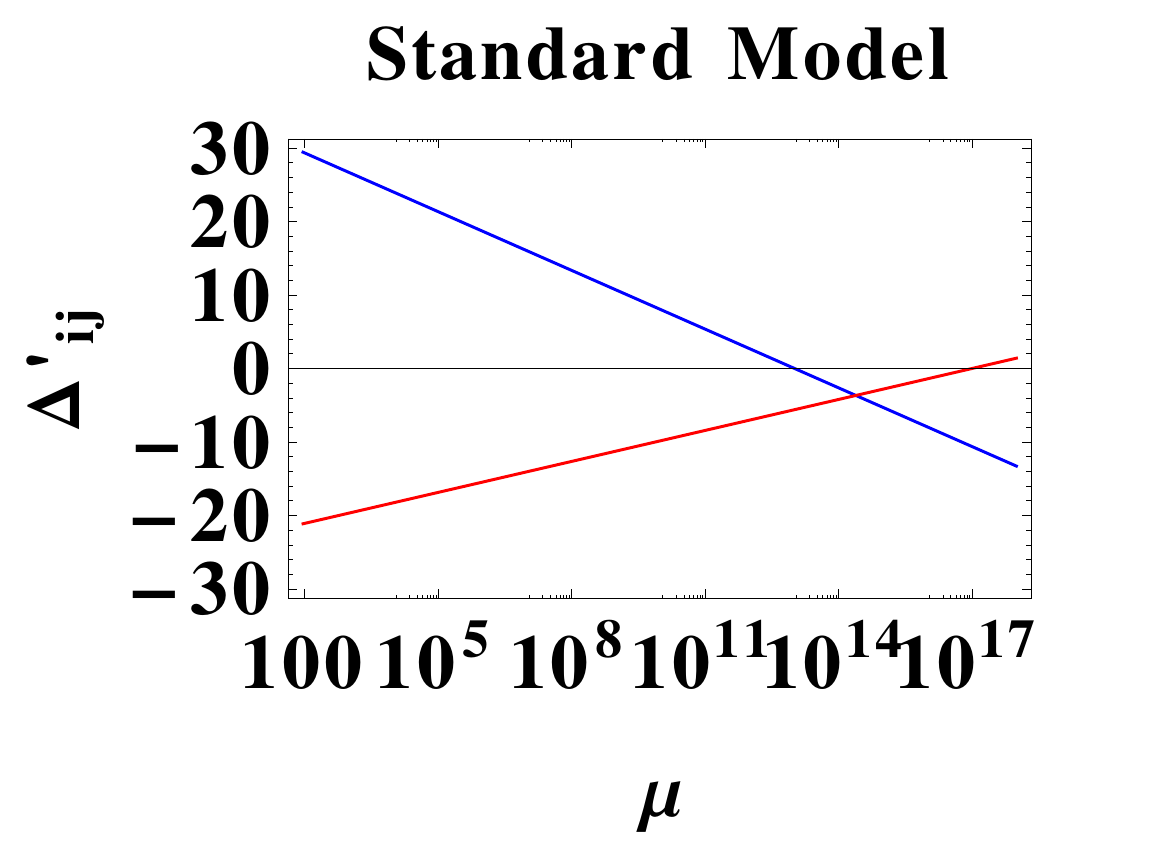}
\includegraphics[bb=0 0 350 242,height=3.5cm]{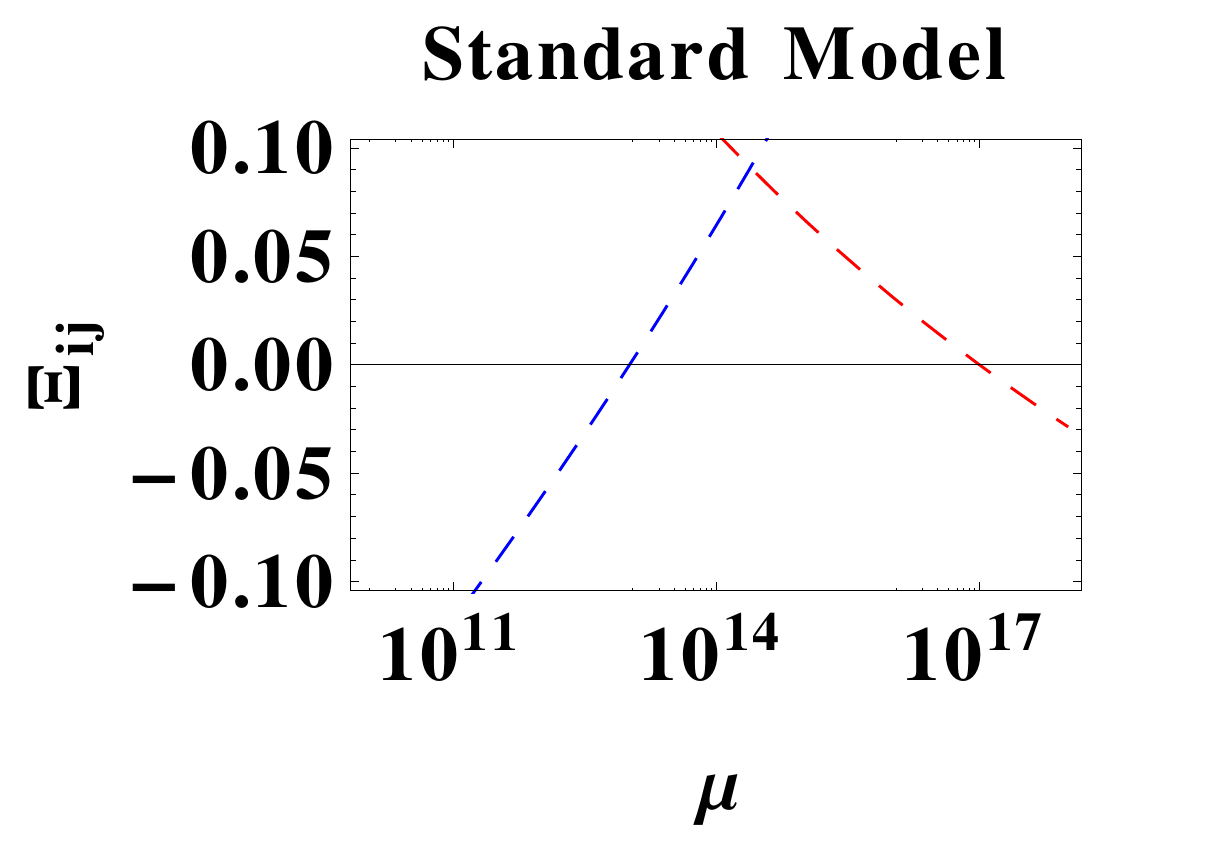}
\end{center}
\caption{
$\mu-\alpha^{-1}(\mu),\mu-\Delta'_{ij}(\mu),\mu-\Xi_{ij}(\mu)$
(Log-Linear plots) in the SM:
the left figure shows $\mu-\alpha^{-1}(\mu)$ (Log-Linear plots), where
the red line is $\alpha_{3C}$,
the green line is $\alpha_{2L}$, and 
the blue line is $\alpha_{1Y}$;
the center figure shows
$\mu-\Delta'_{ij}(\mu)$ (Log-Linear plots), where
the red line is $\Delta'_{3C,2L}=\alpha_{3C}^{-1}-\alpha_{2L}^{-1}$, and 
the blue line is $\Delta'_{1Y,2L}=\alpha_{1Y}^{-1}-\alpha_{2L}^{-1}$;
the right figure shows
$\mu-\Xi_{ij}(\mu)$ (Log-Linear plots), where 
the red line is $\Xi_{3C,2L}=\alpha_{3C}/\alpha_{2L}-1$, and 
the blue line is $\Xi_{1Y,2L}=\alpha_{1Y}/\alpha_{2L}-1$.
}
\label{Figure:RGE-gauge-coupling-SM}
\end{figure}

The RGE evolution for the SM gauge coupling constants in the SM is
shown in Fig.~\ref{Figure:RGE-gauge-coupling-SM}, where we used the
following input parameters for the three SM gauge coupling constants at 
$\mu=M_Z=91.1876\pm 0.0021$ 
given in Ref.~\cite{Agashe:2014kda}
\begin{align}
\alpha_{3C}(M_Z)&=0.1184 \pm 0.0007,\\
\alpha_{2L}(M_Z)&=\frac{\alpha_{em}(M_{Z})}{\sin^2\theta_W(M_Z)},\\
\alpha_{1Y}(M_Z)&=\frac{5\alpha_{em}(M_{Z})}{3\cos^2\theta_W(M_Z)},
\end{align}
where the relations between the EW gauge coupling constants
$\alpha_{2L}(\mu)$ and $\alpha_{1Y}(\mu)$ and the electromagnetic (EM)
gauge coupling constant $\alpha_{em}(\mu)$ and the Weinberg angle
$\theta_W(\mu)$ are given by 
\begin{align}
\alpha_{em}(\mu)&=
\frac{3\alpha_{1Y}(\mu)\alpha_{2L}(\mu)}
{3\alpha_{1Y}(\mu)+5\alpha_{2L}(\mu)},\\
\sin^2\theta_W(\mu)&=
\frac{3\alpha_{1Y}(\mu)}{3\alpha_{1Y}(\mu)+5\alpha_{2L}(\mu)}.
\end{align}
The experimental values of the EM gauge coupling constant
and the Weinberg angle given in Ref.~\cite{Agashe:2014kda} are
\begin{align}
\alpha_{em}^{-1}(M_{Z})&=127.916\pm 0.015,\\
\sin^2\theta_W(M_Z)&=0.23116\pm 0.00013.
\end{align}

As well-known, GUTs based on the $SU(5)$ gauge group and also other
higher rank gauge group without intermediate scales predict the SM gauge
coupling unification at the GUT scale $M_{GUT}$. 
The relations between the SM gauge coupling constants
$\alpha_{i}(\mu)$ are given by
\begin{align}
\alpha_{3C}(M_{GUT})=\alpha_{2L}(M_{GUT})=\alpha_{1Y}(M_{GUT}).
\end{align}
They lead to
\begin{align}
\sin^2\theta_W(M_{GUT})
=\frac{3}{8}.
\end{align}
Obviously, $\sin^2\theta_W(M_{GUT})\not=\sin^2\theta_W(M_{Z})$, 
so we have to take into account the effects for the RGEs for the SM
gauge coupling constants between the EW scale and the GUT scale.

At present the value of $\alpha_{i}(M_Z)$ has roughly 4-digit accuracy
according to Ref.~\cite{Agashe:2014kda}.
Thus, it is meaningless to discuss more than 4-digit accuracy for 
$\Xi_{ij}(\mu)$, We regard ${}^\forall|\Xi_{ij}(\mu)|< 10^{-4}$ as an
almost SM gauge coupling unification scale $M_{GCU}$.
From Fig.~\ref{Figure:RGE-gauge-coupling-SM}, in the SM, for any scale
$\mu$, ${}^\forall|\Xi_{ij}(\mu)|$ cannot be less than $10^{-4}$, and then
in the SM without any correction or only negligible ones, three gauge
coupling constants are not unified.
If there are intermediate symmetry breaking scales between an original
GUT scale and the EW scale, then in general they contribute
non-negligible effect for gauge coupling unification;
it is discussed in e.g., 4D $SO(10)$ GUTs 
\cite{Deshpande:1992au,Deshpande:1992em,Mohapatra2002,Altarelli:2013aqa,Meloni:2014rga}
because one of examples is $G_{GUT}=SO(10)\supset G_{PS}\supset G_{SM}$.
The rank of the original GUT gauge group $G_{GUT}$ must be more
than 4 because the rank of the SM gauge group $G_{SM}$ is 4.
The rank of the $SO(11)$ gauge group is 5, so we will discuss its
intermediate scale effect in the $SO(11)$ GHGUTs.

\subsection{Asymptotic freedom and gauge coupling unification}
\label{Sec:Asymptotic-freedom-Gauge-coupling-unification}

\begin{table}[htb]
\begin{center}
\begin{tabular}{cccc}
\hline
$SO(11)$ Irrep.&$d(G)$&$T(R)$&Type\\\hline
$(10000)$&${\bf 11}$&$1$&R\\
$(00001)$&${\bf 32}$&$4$&PR\\
$(01000)$&${\bf 55}$&$9$&R\\
$(20000)$&${\bf 65}$&$13$&R\\
\hline
\end{tabular}
\caption{Summary for representations of the Lie group $SO(11)$
satisfying a condition $T(R)<(21/8)C_2(SO(11)={\bf 55})=189/8$,
where $SO(11)$ Irrep., $d(G)$, $T(R)$, and Type stand for 
the Dynkin label, the dimension, the Dynkin index, and the type of
of the irreducible representations of $SO(11)$,
respectively. R and PR represent real and pseudo-real
representations of $SO(11)$. (See Ref.~\cite{Yamatsu:2015gut} in detail.)
}
\label{Table:Summary-representations-SO(11)}
\end{center}
\end{table}

We check the asymptotic freedom condition given in 
Eq.~(\ref{Eq:Condition-asymptotic-free}) in $SO(11)$ GHGUTs.
To keep the success of the $SO(11)$ gauge-Higgs grand
unification in Ref.~\cite{Hosotani:2015hoa}, such as automatic chiral
anomaly cancellation for the gauge symmetries on the Planck and TeV
branes, we use the same orbifold boundary conditions (BC); the orbifold
BC on the Planck brane $y=0$ breaks $SO(11)$ to $SO(10)$; the orbifold
BC on the TeV brane $y=L$ breaks $SO(11)$ to 
$SO(4)\times SO(7)\simeq SU(2)\times SU(2)\times SO(7)$. 
The two orbifold BCs break $SO(11)$ to the Pati-Salam gauge
group $G_{PS}$.
The orbifold boundary conditions for the $SO(11)$ vector
representation ${\bf 11}$ on the Planck and TeV branes are given by 
\begin{align}
P_{0{\bf 11}}=\mbox{diag}\left(I_{10},-I_1\right),\ \ \
P_{1{\bf 11}}=\mbox{diag}\left(I_{4},-I_7\right).
\end{align}
Also, by using the branching rules of the representations in 
Table~\ref{Table:Summary-representations-SO(11)} shown in 
Ref.~\cite{Yamatsu:2015gut}, we find that 
the branching rules of pseudo-real representations of $SO(11)$ lead to 
complex representations of its subgroup, while the branching rules of
real representations of $SO(11)$ lead to real representations of its
subgroup. That is, we must use pseudo-real representations to realize a 4D
chiral gauge theory. In Table~\ref{Table:Summary-representations-SO(11)},
only the $SO(11)$ spinor representation ${\bf 32}$ is a pseudo-real
representation of $SO(11)$. (The $SO(11)$ ${\bf 320}$ representation is 
the second lowest dimensional pseudo-real representation listed in
Ref.~\cite{Yamatsu:2015gut}.)
Also, the zero modes of each $SO(11)$ spinor bulk fermion field are the
five SM fermions plus one right-hand neutrino. 
Therefore, the matter content of $SO(11)$ GHGUTs must contain at
least three $SO(11)$ spinor bulk fermion fields as the same as that
in Ref.~\cite{Hosotani:2015hoa}, so we subtract the contribution from
the three $SO(11)$ spinor bulk fermion fields.
The asymptotic freedom condition is
\begin{align}
\sum_{R}T(R)<\frac{93}{8}.
\label{Eq:Condition-asymptotic-free-SO11}
\end{align}

We consider which matter contents can satisfy the asymptotic freedom
condition in Eq.~(\ref{Eq:Condition-asymptotic-free-SO11}).
To maintain the number of chiral matter fields, if we introduce an 
$SO(11)$ spinor bulk fermion field with a parity assignment, then
we must also introduce another $SO(11)$ spinor bulk fermion field with a
the opposite parity assignment. 
From Table~\ref{Table:Summary-representations-SO(11)}, the $SO(11)$
${\bf 65}$ representation does not satisfy the condition.
By using the condition in Eq.~(\ref{Eq:Condition-asymptotic-free-SO11})
and the Dynkin indices given in Table~\ref{tab:matter-content}, 
we summarize the matter contents in
Table~\ref{Table:three-chiral-generations-asymptotic-free-SO11} that 
satisfy three chiral generations of quarks and leptons and the
asymptotic freedom condition in
Eq.~(\ref{Eq:Condition-asymptotic-free-SO11}). 

\begin{table}[htb]
\begin{center}
\begin{tabular}{cccc}
\hline
$n_{\bf 55}$&$n_{\bf 32}$&$n_{\bf 11}$&$\Delta b^{KK}$\\
\hline
$0$&$3$&$0$      &$-\frac{31}{2}$\\
$0$&$3$&$\leq 11$&$\frac{-93+8n_{\bf 11}}{6}$\\
$0$&$5$&$\leq 3$ &$\frac{-29+8n_{\bf 11}}{6}$\\
$1$&$3$&$\leq 2$ &$\frac{-21+8n_{\bf 11}}{6}$\\
\hline
\end{tabular}
\caption{Matter contents that satisfy three chiral
 generations of quarks and leptons and the asymptotic freedom condition in
Eq.~(\ref{Eq:Condition-asymptotic-free-SO11}). 
}
\label{Table:three-chiral-generations-asymptotic-free-SO11}
\end{center}
\end{table}

In the $SO(11)$ GHGUT \cite{Hosotani:2015hoa}, a fermion number
conservation lead to sufficient proton decay suppression
\cite{Hosotani:2015hoa}.  
When we impose the fermion number conservation,
an $SO(11)$ ${\bf 55}$ bulk fermion with an orbifold BCs must have
another $SO(11)$ ${\bf 55}$ bulk fermion with the opposite orbifold BCs; 
an $SO(11)$ ${\bf 11}$ bulk fermion with a orbifold BCs must have
another  $SO(11)$ ${\bf 11}$ bulk fermion with the opposite orbifold
BCs. From Table~\ref{Table:three-chiral-generations-asymptotic-free-SO11},
we cannot introduce any $SO(11)$ ${\bf 55}$ bulk fermion to keep the
fermion number conservation without exotic fermion zero modes.
The matter contents that satisfy three chiral  generations of
quarks and leptons, the asymptotic freedom condition in
Eq.~(\ref{Eq:Condition-asymptotic-free-SO11}), and the fermion number
conservation are shown in
Table~\ref{Table:three-chiral-generations-asymptotic-free-fermion-number-SO11}.

\begin{table}[htb]
\begin{center}
\begin{tabular}{cccc}
\hline
$n_{\bf 32}$&$n_{\bf 11}$&$\Delta b^{KK}$\\
\hline
$3$&$0$      &$-\frac{31}{2}$\\
$5$&$\leq 2$      &$\frac{-13}{6}$\\
$3$&$\leq 10$&$\frac{-93+8n_{\bf 10}}{6}$\\
\hline
\end{tabular}
\caption{Matter contents that satisfy three chiral
 generations of quarks and leptons, the asymptotic freedom condition in
Eq.~(\ref{Eq:Condition-asymptotic-free-SO11}),
and the fermion number conservation. 
}
\label{Table:three-chiral-generations-asymptotic-free-fermion-number-SO11}
\end{center}
\end{table}

\begin{figure}[tbh]
\begin{center}
\includegraphics[bb=0 0 357 261,height=3.5cm]{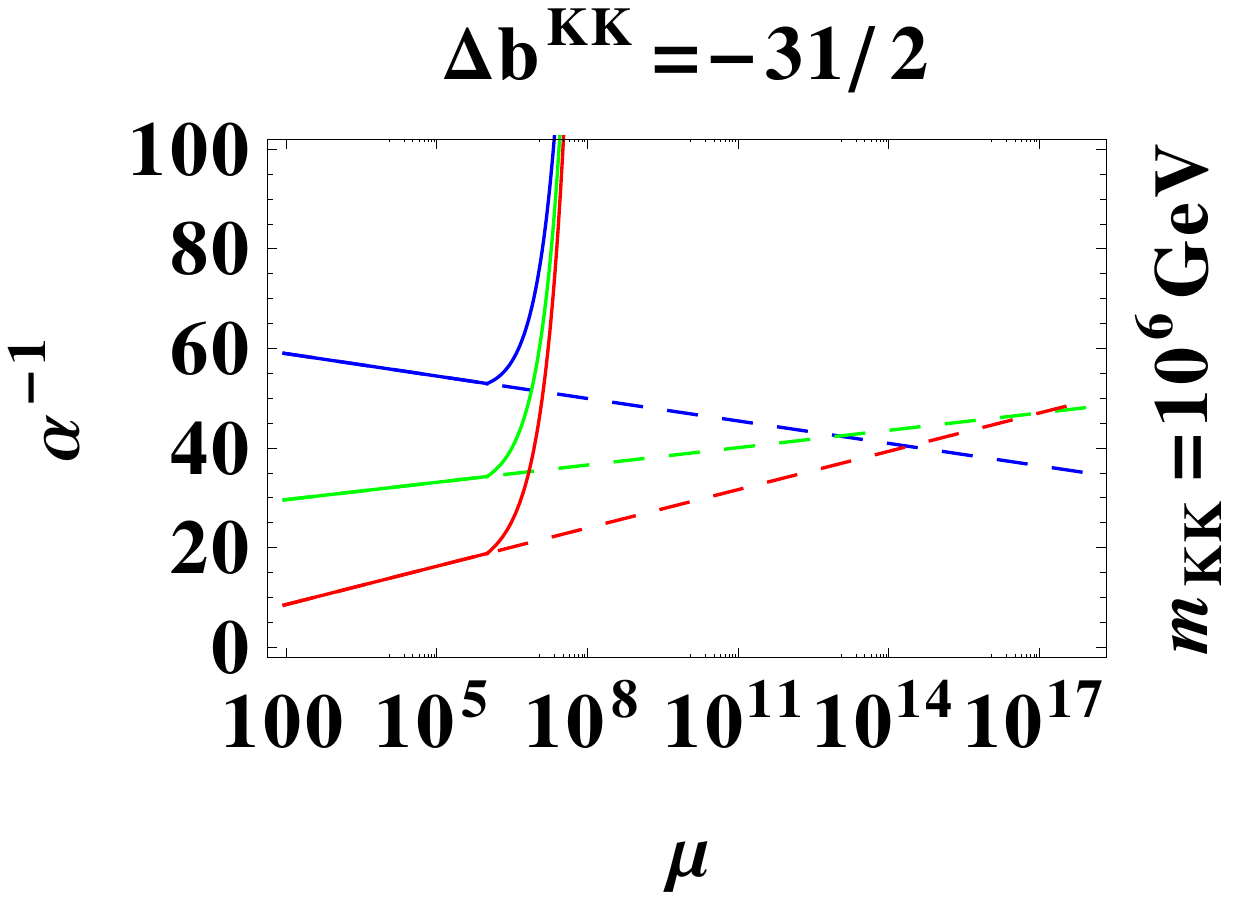}
\includegraphics[bb=0 0 358 262,height=3.5cm]{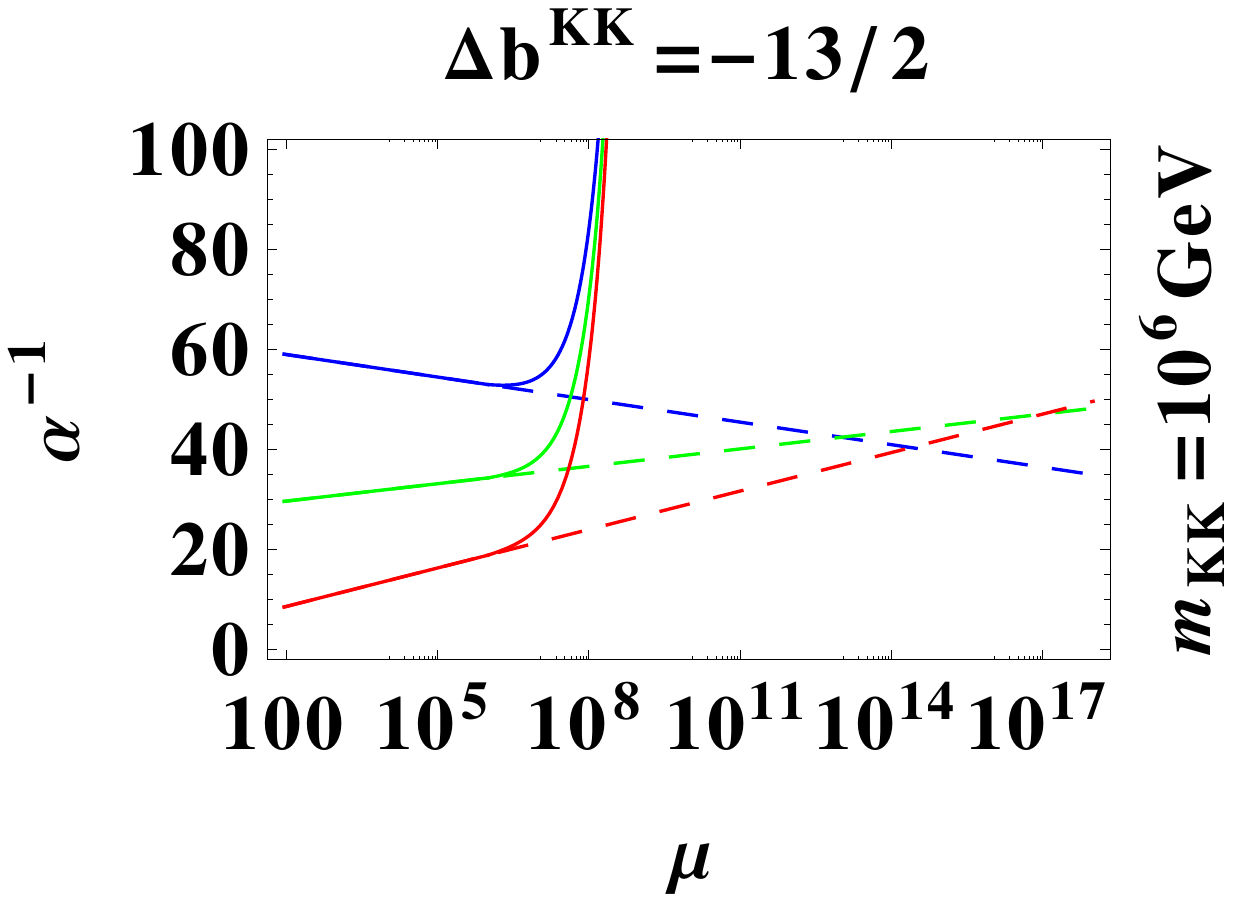}
\includegraphics[bb=0 0 356 261,height=3.5cm]{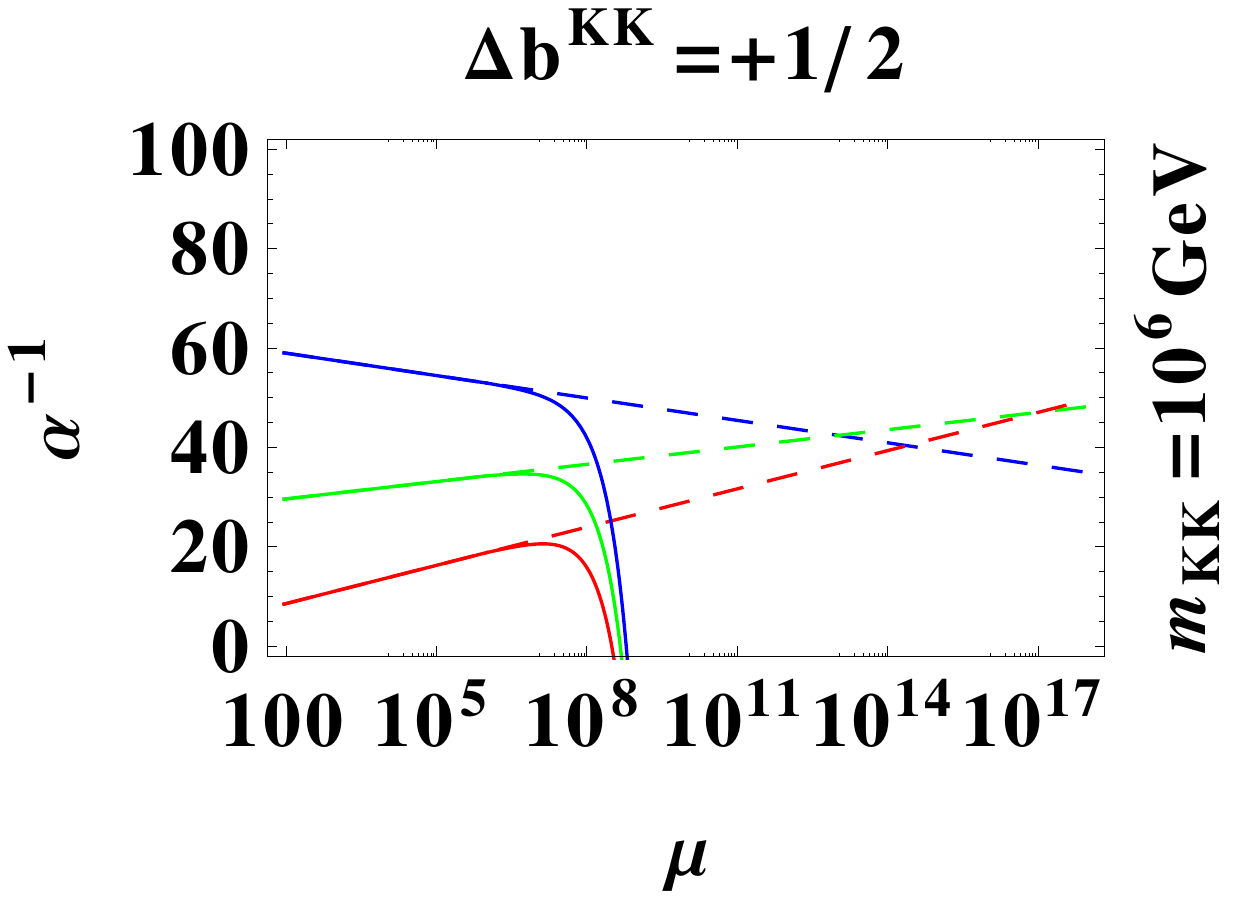}
\end{center}
\caption{$\mu-\alpha^{-1}(\mu)$ (Log-Linear plots) in $SO(11)$ GHGUTs
 with one KK mass scale $m_{KK}=10^{10}$ GeV:
the left, center, and right figures show 
$(n_{\bf 32},n_{\bf 11})=(3,0)$ ($\Delta b^{KK}=-31/2$),
$(n_{\bf 32},n_{\bf 11})=(5,2)$ ($\Delta b^{KK}=-13/6$), and 
$(n_{\bf 32},n_{\bf 11})=(5,4)$ ($\Delta b^{KK}=+1/2$), respectively,
where the real lines show the $SO(11)$ GHGUTs;
the dashed lines show the SM ones;
the red lines stand for $\alpha_{3C}$;
the green lines stand for $\alpha_{2L}$;
the blue lines stand for $\alpha_{1Y}$.
}
\label{Figure:RGE-gauge-coupling-GHGU-vs-SM}
\end{figure}

\begin{figure}[tbh]
\begin{center}
\includegraphics[bb=0 0 435 303,width=7.5cm]{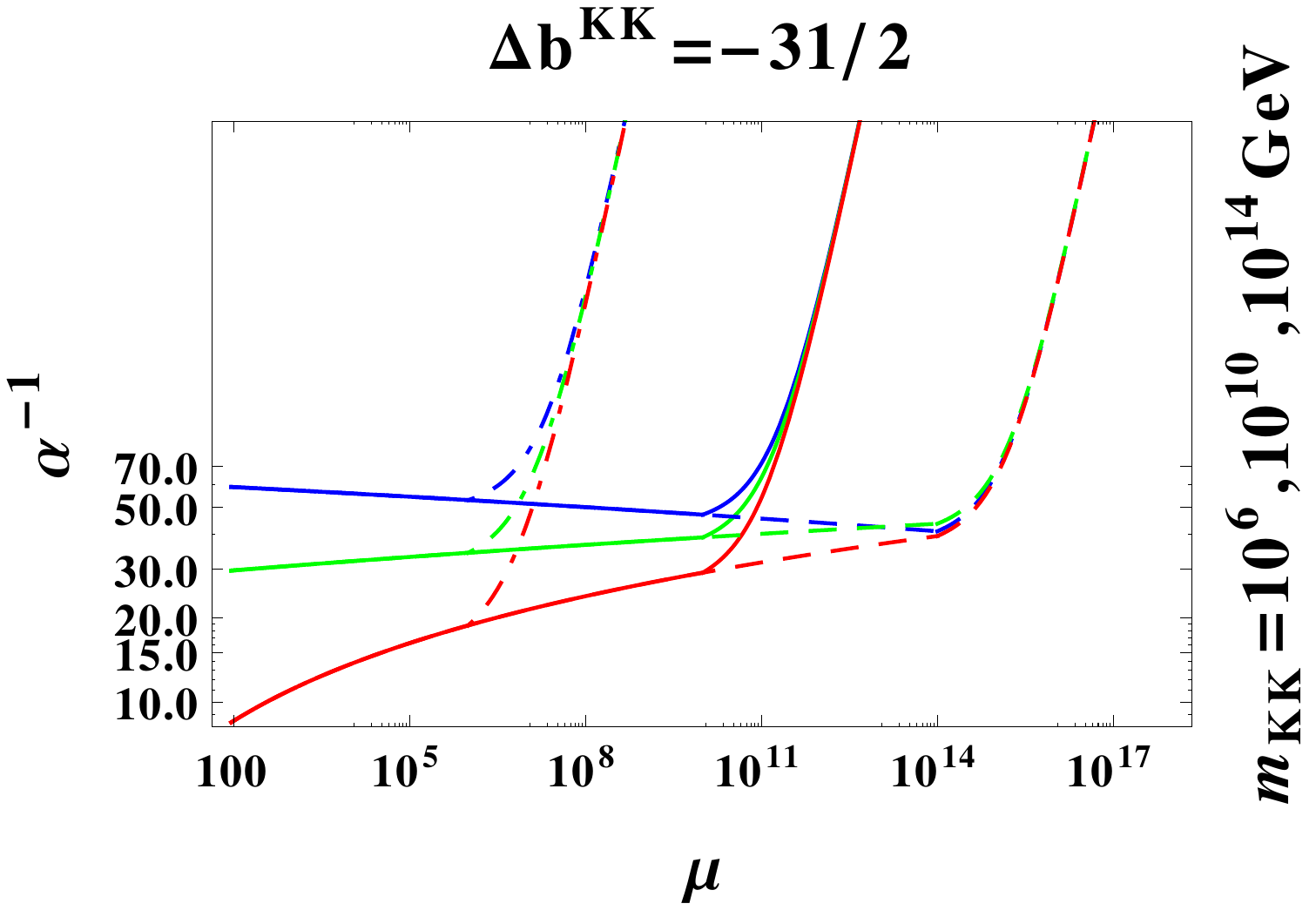}
\includegraphics[bb=0 0 434 303,width=7.5cm]{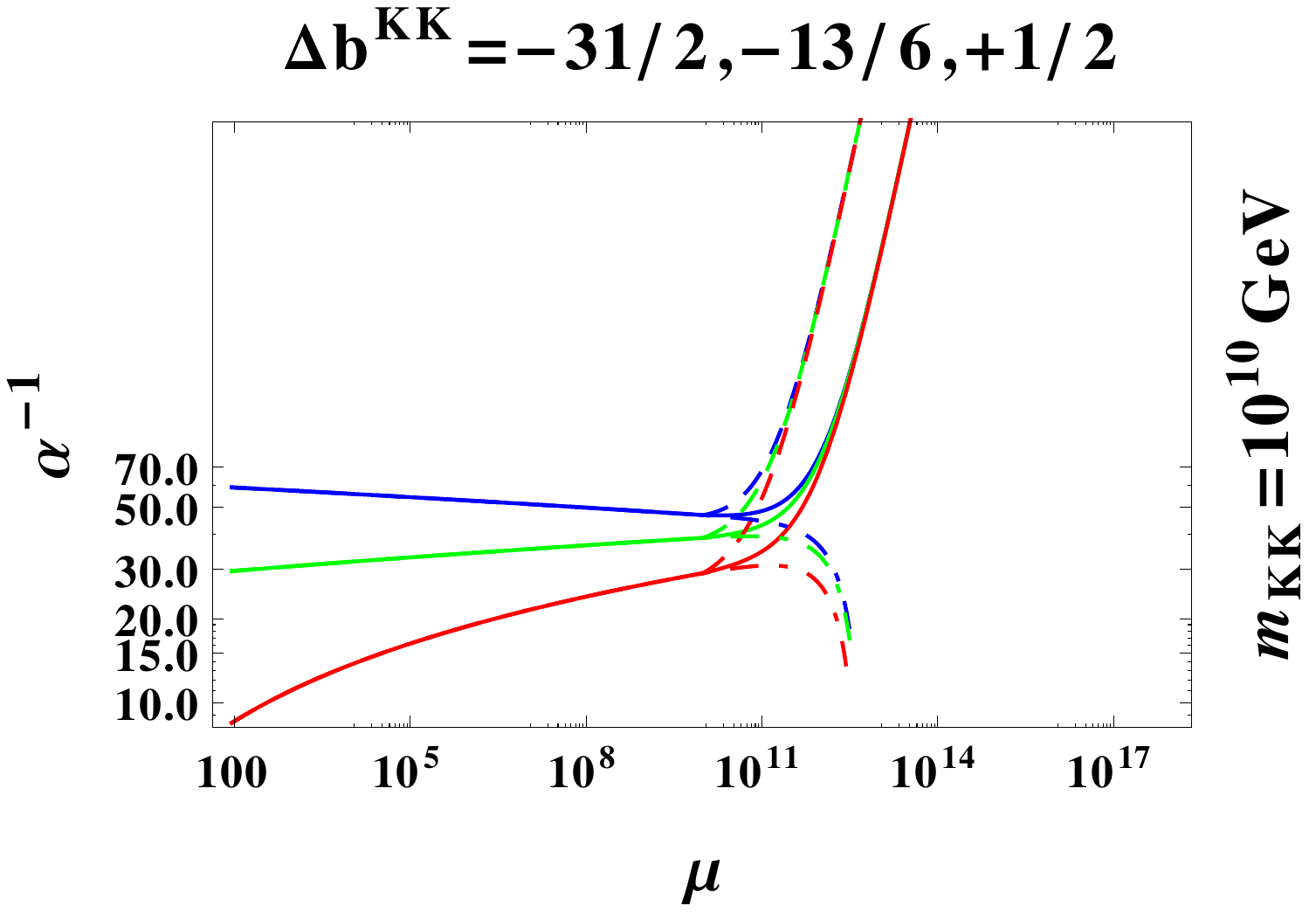}
\end{center}
\caption{$\mu-\alpha^{-1}(\mu)$ (Log-Log plots) in $SO(11)$ GHGUTs:
the left figure shows three different matter contents
$(n_{\bf 32},n_{\bf 11})=(3,0)$ ($\Delta b^{KK}=-31/2$),
$(n_{\bf 32},n_{\bf 11})=(5,2)$ ($\Delta b^{KK}=-13/6$), 
$(n_{\bf 32},n_{\bf 11})=(5,4)$ ($\Delta b^{KK}=+1/2$), 
with a fixed KK mass $m_{KK}=10^{10}$ GeV,
where 
the dashed lines show $\Delta b^{KK}=-31/2$,
the real lines show $\Delta b^{KK}=-13/6$, and 
the dash-doted lines show $\Delta b^{KK}=+1/2$;
the left figure shows one matter content
$(n_{\bf 32},n_{\bf 11})=(3,0)$ ($\Delta b^{KK}=-31/2$),
with three different KK masses $m_{KK}=10^{6},10^{10},10^{14}$ GeV,
where 
the dashed lines show $m_{KK}=10^6$ GeV,
the real lines show $m_{KK}=10^{10}$ GeV, and 
the dash-doted lines show $m_{KK}=10^{14}$ GeV.
For all the figures, the red lines stand for $\alpha_{3C}$;
the green lines stand for $\alpha_{2L}$;
the blue lines stand for $\alpha_{1Y}$.
}
\label{Figure:RGE-gauge-coupling-GHGU-Alpha}
\end{figure}

As in the previous section, we use approximate mass spectra of zero
modes and $k$-th KK modes whose masses are $m=0$ and $m=km_{KK}$,
respectively. We also use the gauge coupling constant in
Eq.~(\ref{Eq:4D-gauge-coupling-constant-in-5D}) for the three SM gauge 
group, where $\alpha^{-1}$, $b^0$, and $\Delta b^{KK}$ should be
replaced by $\alpha_{i}^{-1}$, $b_i^0$, and $\Delta b^{KK}$.
$\alpha_{i}^{-1}$ and $b_i^0$ are dependent on the SM gauge group, 
while $\Delta b^{KK}$ is independent from the SM gauge group.
From Eq.~(\ref{Eq:4D-gauge-coupling-constant-in-5D}), we find that the
difference between the $SO(11)$ GHGUTs and the SM is only its third term
dependent on $\Delta b^{KK}$ for $\mu>m_{KK}$.
Also, the difference between $\alpha_{i}^{-1}$ and 
$\alpha_{j}^{-1}$ $(i\not=j)$ in the $SO(11)$ GHGUTs is the first and
second terms in Eq.~(\ref{Eq:4D-gauge-coupling-constant-in-5D}).
Therefore, $\Delta'_{ij}(\mu)$ in the $SO(11)$ GHGUTs
are the same as those in the SM. 
($\Delta_{ij}'(\mu)$ in the SM are shown in the center figure in
Fig.~\ref{Figure:RGE-gauge-coupling-SM}.)
By using the asymptotic form of the gauge coupling constant given in
Eq.~(\ref{Eq:gauge-coupling-convergence}), for $\mu\gg m_{KK}$, 
$\Xi_{ij}(\mu)$ can be written as 
\begin{align}
\Xi_{ij}(\mu)\simeq
-\Delta_{ij}'(\mu)
\left(\frac{-2\pi}{\Delta b^{SO(11)}}\frac{m_{KK}}{\mu}\right).
\label{Eq:Xi-Delta-relation}
\end{align}

\begin{figure}[tbh]
\begin{center}
\includegraphics[bb=0 0 357 261,height=3.5cm]{RGE-gauge-SO11_b=-31_mKK-6_Alpha.pdf}
\includegraphics[bb=0 0 336 245,height=3.5cm]{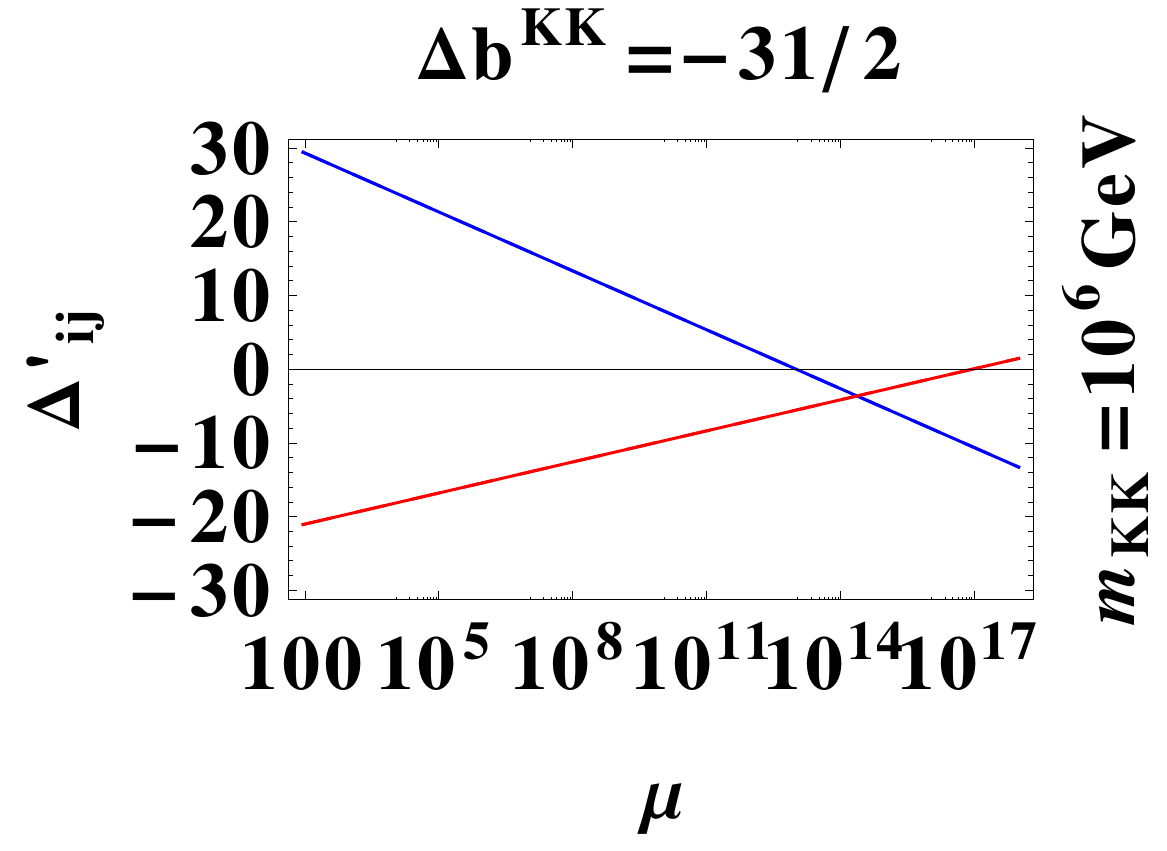}
\includegraphics[bb=0 0 391 245,height=3.5cm]{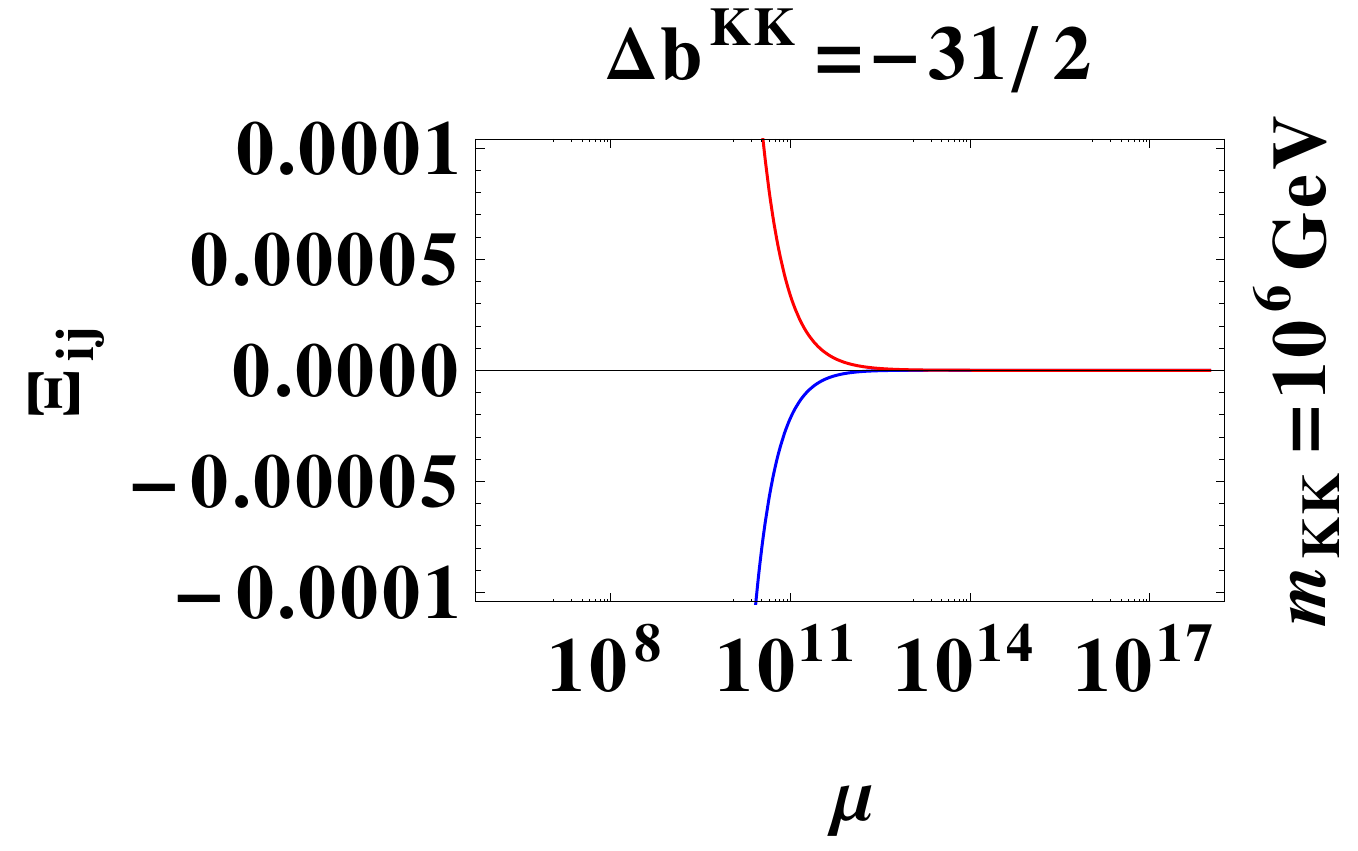}
\end{center}
\caption{$\mu-\alpha^{-1}(\mu)$, $\mu-\Delta'_{ij}(\mu)$,
$\mu-\Xi_{ij}(\mu)$ (Log-Linear plots) in $SO(11)$ GHGUT with
the same matter content 
$(n_{\bf 32},n_{\bf 11})=(3,0)$ ($\Delta b^{KK}=-31/2$)
and KK mass $m_{KK}=10^{6}$ GeV:
the left figures show $\mu-\alpha^{-1}(\mu)$ (Log-Linear plots), where
the red line represents $\alpha_{3C}$,
the green line represents $\alpha_{2L}$, and 
the blue line represents $\alpha_{1Y}$;
the center figures show
$\mu-\Delta'_{ij}(\mu)$ (Log-Linear plots), where
the red line is $\Delta'_{3C,2L}=\alpha_{3C}^{-1}-\alpha_{2L}^{-1}$, and 
the blue line is $\Delta'_{1Y,2L}=\alpha_{1Y}^{-1}-\alpha_{2L}^{-1}$;
the right figures show
$\mu-\Xi_{ij}(\mu)$ (Log-Linear plots), where 
the red line is $\Xi_{3C,2L}=\alpha_{3C}/\alpha_{2L}-1$, and 
the blue line is $\Xi_{1Y,2L}=\alpha_{1Y}/\alpha_{2L}-1$.
The dashed lines show the SM,
the real lines show $SO(11)$ GHGUT.
}
\label{Figure:RGE-gauge-coupling-GHGU}
\end{figure}

Let us check what we can learn from 
Figs.~\ref{Figure:RGE-gauge-coupling-GHGU-vs-SM},
\ref{Figure:RGE-gauge-coupling-GHGU-Alpha}, and
~\ref{Figure:RGE-gauge-coupling-GHGU}.
From Fig.~\ref{Figure:RGE-gauge-coupling-GHGU-vs-SM},
we can clearly see that the three SM gauge coupling constants 
$\alpha_i (i=3C,2L,1Y)$ are
convergent into one and rapidly decreasing for 
$\Delta b^{KK}<0$ and increasing for 
$\Delta b^{KK}>0$ as shown in
Eq.~(\ref{Eq:gauge-coupling-convergence}) 
and Eq.~(\ref{Eq:gauge-coupling-divergence}), respectively.
From the left figure in Fig.~\ref{Figure:RGE-gauge-coupling-GHGU-Alpha},
we find that for $\Delta b^{KK}<0$, the SM gauge coupling
constants decrease drastically above $m_{KK}$ and also they converge,
where the convergent scales depend on $\Delta b^{KK}$;
for $\Delta b^{KK}>0$, the SM gauge coupling constants increase
drastically and also seem to converge above $m_{KK}$, where our
perturbative calculation is not reliable.
From the right figure in Fig.~\ref{Figure:RGE-gauge-coupling-GHGU-Alpha}, 
we find that regardless of KK mass scales
$m_{KK}=10^{6},10^{10},10^{14}$ GeV,  
the SM gauge coupling constants decrease drastically above $m_{KK}$ and
also they converge for $\Delta b^{KK}<0$, where they diverge 
for $\Delta b^{KK}>0$.
From the right figure in Fig.~\ref{Figure:RGE-gauge-coupling-GHGU}, we
find that for $\mu\sim 10^{10.5}$ GeV, $\Xi_{ij}(\mu)\sim 10^{-4}$, so we
regard $\mu>10^{10.5}$ GeV as $M_{GCU}$ in this case.
The center figure in
Fig.~\ref{Figure:RGE-gauge-coupling-GHGU} is exactly the same as 
that in Fig.~\ref{Figure:RGE-gauge-coupling-SM}. One may wonder that 
even in the $SO(11)$ GHGUTs, the gauge coupling constants were not unified 
based on Fig.~\ref{Figure:RGE-gauge-coupling-GHGU}.
For the $SO(11)$ GHGUTs, from the definition of $\Delta'_{ij}(\mu)$ in 
Eq.~(\ref{Eq:difference-gauge-coupling-2}) and 
the 4-digit accuracy of $\alpha_i(M_Z)$, the error of 
$\Delta'_{ij}(\mu)$ is 
$\mbox{Err}[\Delta'_{ij}(\mu)]\simeq
O(10^{-4})\alpha^{-1}(\mu)\simeq O(10^{-4})\mu/m_{KK}$
for $\mu>M_{GCU}\gg m_{KK}$.
For $m_{KK}=10^{6}$ GeV and $\mu=10^{11}$ GeV,
$\mbox{Err}[\Delta'_{ij}(10^{11}\mbox{GeV})]\simeq O(10)$ and
the deviations $\Delta'_{3C,2L}(10^{11}\mbox{GeV})$ 
and $\Delta'_{1Y,2L}(10^{11}\mbox{GeV})$ are less than 10 from the
center figure in Fig.~\ref{Figure:RGE-gauge-coupling-GHGU}, and then
$M_{GCU}$ starts around $10^{11}$ GeV.

From the above discussion, we found that in the $SO(11)$ GHGUTs
the 4D SM gauge coupling constants are almost unified above
$M_{GCU}$ regardless of the matter contents and their mass spectra
and the SM gauge coupling constants are asymptotically free.
We also found that $M_{GCU}$ depends on the matter contents given in
Table~\ref{Table:three-chiral-generations-asymptotic-free-SO11}. 

\subsection{Corrections for gauge coupling constants}
\label{Sec:Corrections}

We check whether the above analysis is valid even when we take into
account several corrections. We divide our discussion into two cases,
$m_{KK}<M_{PS}\simeq M_{GUT}=1/L$ and $m_{KK}<M_{PS}<M_{GUT}=1/L$, where
$M_{PS}$ is the symmetry breaking scale at which $G_{PS}$ gauge symmetry
is broken in $G_{SM}$ gauge symmetry. This is because for
$m_{KK}<M_{GUT}=1/L\simeq M_{PS}$, we use only the RGEs for the $G_{SM}$
gauge coupling constants, while for $m_{KK}<M_{PS}<M_{GUT}=1/L$,
we have to use the RGEs for the $G_{SM}$ gauge coupling constants below
$M_{PS}$ and the RGEs for the $G_{PS}$ gauge coupling constants above
$M_{PS}$. In the latter analysis, we have to take int account the
matching conditions between the $G_{PS}$ gauge coupling constraints 
and the $G_{SM}$ gauge coupling constraints at the Pati-Salam scale
$M_{PS}$. (Note that for 4D non-SUSY $SO(10)$ GUTs, this effect has been 
discussed in many articles, e.g.,
Refs.~\cite{Deshpande:1992au,Deshpande:1992em,Mohapatra2002,Altarelli:2013aqa,Meloni:2014rga,Mambrini:2015vna,Babu:2015bna}.)

\subsubsection{$m_{KK}<M_{GUT}\simeq M_{PS}\simeq 1/L$}

Here we check whether the above analysis is valid even when we take into
account mass spectra of bulk fields.
Since mass spectra in the $SO(11)$ GHGUTs depend on orbifold BCs and
parameters of bulk and brane terms, it is almost 
impossible to use them in exact expression. Instead of them, we use
approximate forms for flat space limit. We use the mass spectra of $k$th
KK modes $(k=1,2,\cdots)$ of bulk fields by their orbifold BCs for flat
space limit: 
\begin{align}
(N,D),\ (D,N):\ &\frac{2k-1}{2}m_{KK},\\
(N,N),\ (D,D):\ &k m_{KK},
\end{align}
where $N$ and $D$ stand for Neumann and Dirichlet BCs, respectively. 
$(X,Y)$ $(X,Y=N,D)$ stands for the orbifold BCs on the Planck and TeV
branes, respectively. (This approximation is good for large $k$ because 
the RS warped space is asymptotically flat space for short distance.)
Only each field with $(N,N)$ contains a zero mode. For large $k$, a
$k$th KK mass spectrum in RS warped space is approaching to 
that in flat space. For almost cases, the difference between warped and
flat spaces leads to only tiny effect for RGEs because the contribution
to the $\beta$-function coefficient from each mode is logarithmic.
In the following discussion, we use the above approximate mass spectra.

By using the above approximation about mass spectra of the bulk fields,
the RGE for the gauge coupling constant can be divided into three
regions: 
\begin{align}
\frac{d}{d\mbox{log}(\mu)}\alpha_i^{-1}\simeq
\left\{
\begin{array}{ll}
-\frac{1}{2\pi}b_i^{0}&\ \ \mbox{for}\ \ \mu<\frac{m_{KK}}{2}\\
-\frac{1}{2\pi}
\left(b_i^0+\delta b_i^{\rm KK}+(k-1)\Delta b^{\rm KK}\right)
&\ \ \mbox{for}\ \
\left(k-\frac{1}{2}\right)m_{KK}\leq\mu<km_{KK}\\
-\frac{1}{2\pi}
\left(b_i^0+k\Delta b^{\rm KK}\right)&\ \ \mbox{for}\ \
km_{KK}\leq\mu<\left(k+\frac{1}{2}\right)\\
\end{array}
\right.,
\label{Eq:RGE-4D-gauge-coupling-in-5D-f}
\end{align}
where $b_i^0$ is a $\beta$-function coefficients given from its zero
modes, i.e., bulk fields with the orbifold BC $(N,N)$; 
$\delta b_i^{\rm KK}$ is an $\beta$-function coefficient by  
bulk fields with the orbifold BC $(N,D)$ or $(D,N)$; 
$\Delta b^{\rm KK}$ is an additional $\beta$-function coefficient
generated by a set of KK modes of all bulk fields, where 
$b_i^0$, $\delta b_i^{\rm KK}$, and $\Delta b^{\rm KK}$ can be 
calculated by using Eq.~(\ref{Eq:beta-function-coeff-general}).

We solve the RGE in Eq.~(\ref{Eq:RGE-4D-gauge-coupling-in-5D-f}).
As in Sec.~\ref{Sec:General}, the number of the set of KK modes for
$\mu> m_{KK}$ is approximately equal to the energy scale divided by the
KK mass scale $k\simeq\mu/m_{KK}$ in Eq.~(\ref{Eq:Approximation}). 
Under the approximation, we can solve the RGE, exactly, but that seems
to be hard to see the contribution from mass splitting effects. We only
write down rough approximate form for $m_{KK}\geq\mu$, 
\begin{align}
\alpha_i^{-1}(\mu)\simeq&
\alpha_i^{-1}(m_{KK})
-\left(\frac{b_i^0}{2\pi}+\frac{\delta b_i^{KK}}{4\pi}\right)
\mbox{log}\left(\frac{\mu}{m_{KK}}\right)
-\frac{\Delta b^{\rm KK}}{2\pi}
\left(\frac{\mu}{m_{KK}}-1\right),
\label{Eq:4D-gauge-coupling-constant-asymptotic-in-5D-f}
\end{align}
where for $M_Z<\mu<m_{KK}$,
\begin{align}
\alpha_i^{-1}(\mu)&=
\alpha_i^{-1}(M_Z)
-\frac{1}{2\pi}b_i^0
\mbox{log}\left(\frac{\mu}{M_Z}\right).
\end{align}
(For the above expression, we ignored the contribution to
$\alpha_i(\mu)$ from $\delta b_i^{KK}$ between $m_{KK}/2$ and $m_{KK}$,
and etc.)
We find that the first and second terms in
Eq.~(\ref{Eq:4D-gauge-coupling-constant-asymptotic-in-5D-f})
are negligible compared with the third term for large $\mu$.

\begin{table}[htb]
\begin{center}
\begin{tabular}{ccl}
\hline
Field&BC&Representations of $G_{SM}=SU(3)_C\times SU(2)_L\times U(1)_Y$\\
\hline
$A_\mu$&
 $(N,N)$&$({\bf 8,1})_0$, $({\bf 1,1})_0$, $({\bf 1,3})_0$\\
&$(N,D)$&$({\bf 3,2})_{-5/6}$, $({\bf \overline{3},2})_{+5/6}$\\
&$(D,N)$&$({\bf 3,1})_{-1/3}$, $({\bf \overline{3},1})_{+1/3}$\\
&$(D_{\rm eff},N)$&$({\bf 3,1})_{+2/3}$, $({\bf \overline{3},1})_{-2/3}$, 
         $({\bf 1,1})_{+1}$, $({\bf 1,1})_{-1}$, $({\bf 1,1})_{0}$\\
&$(D,D)$&$({\bf 1,2})_{+1/2}$, $({\bf 2,1})_{-1/2}$\\
&$(D_{\rm eff},D)$&$({\bf 3,2})_{+1/6}$, $({\bf \overline{3},2})_{-1/6}$\\
$A_y$&
 $(N,N)$&$({\bf 1,2})_{+1/2}$, $({\bf 2,1})_{-1/2}$\\
&$(N,D)$&$({\bf 3,1})_{-1/3}$, $({\bf \overline{3},1})_{+1/3}$\\
&$(D,N)$&$({\bf 3,2})_{-5/6}$, $({\bf \overline{3},2})_{+5/6}$, 
         $({\bf 3,2})_{+1/6}$, $({\bf \overline{3},2})_{-1/6}$\\
&$(D,D)$&$({\bf 8,1})_0$, $({\bf 1,1})_0$, $({\bf 1,3})_0$,
         $({\bf 3,1})_{+2/3}$, $({\bf \overline{3},1})_{-2/3}$, 
         $({\bf 1,1})_{+1}$, $({\bf 1,1})_{-1}$, $({\bf 1,1})_{0}$\\
\hline
\end{tabular}
\caption{The orbifold BCs of the components of the $SO(11)$
 bulk gauge field $A_M=A_\mu\oplus A_y$
}
\label{Table:SO(11)-bulk-gauge-field-BCs}
\end{center}
\end{table}

By using the above discussion, we calculate how much the mass splitting
effect by the orbifold BCs contributes to the gauge coupling unification.
First, we need to know the contribution for $\delta b_i^{KK}$ from
the $SO(11)$ bulk gauge fields and the $SO(11)$ ${\bf 32}$ and 
${\bf 11}$ bulk fermion fields, but as long as the fermion number is
preserved and their brane Dirac mass terms change the component fields
with a Neumann BC to those with an effective Dirichlet BC
on the Planck brane, they lead to the same
contribution to all three gauge coupling constants: 
$\delta b_{1L}^{KK}=\delta b_{2L}^{KK}=\delta b_{3C}^{KK}$.
Therefore, we consider the contribution to $\delta b_i^{KK}$
$(i=3C,2L,1Y)$ from the $SO(11)$ bulk gauge field.
From Table~\ref{Table:SO(11)-bulk-gauge-field-BCs}, we get
\begin{align}
\delta b_{3C}^{KK}=-\frac{83}{6},\ \
\delta b_{2L}^{KK}=-10,\ \
\delta b_{1Y}^{KK}=-\frac{437}{15}.
\end{align}

\begin{figure}[tbh]
\begin{center}
\includegraphics[bb=0 0 357 261,height=3.5cm]{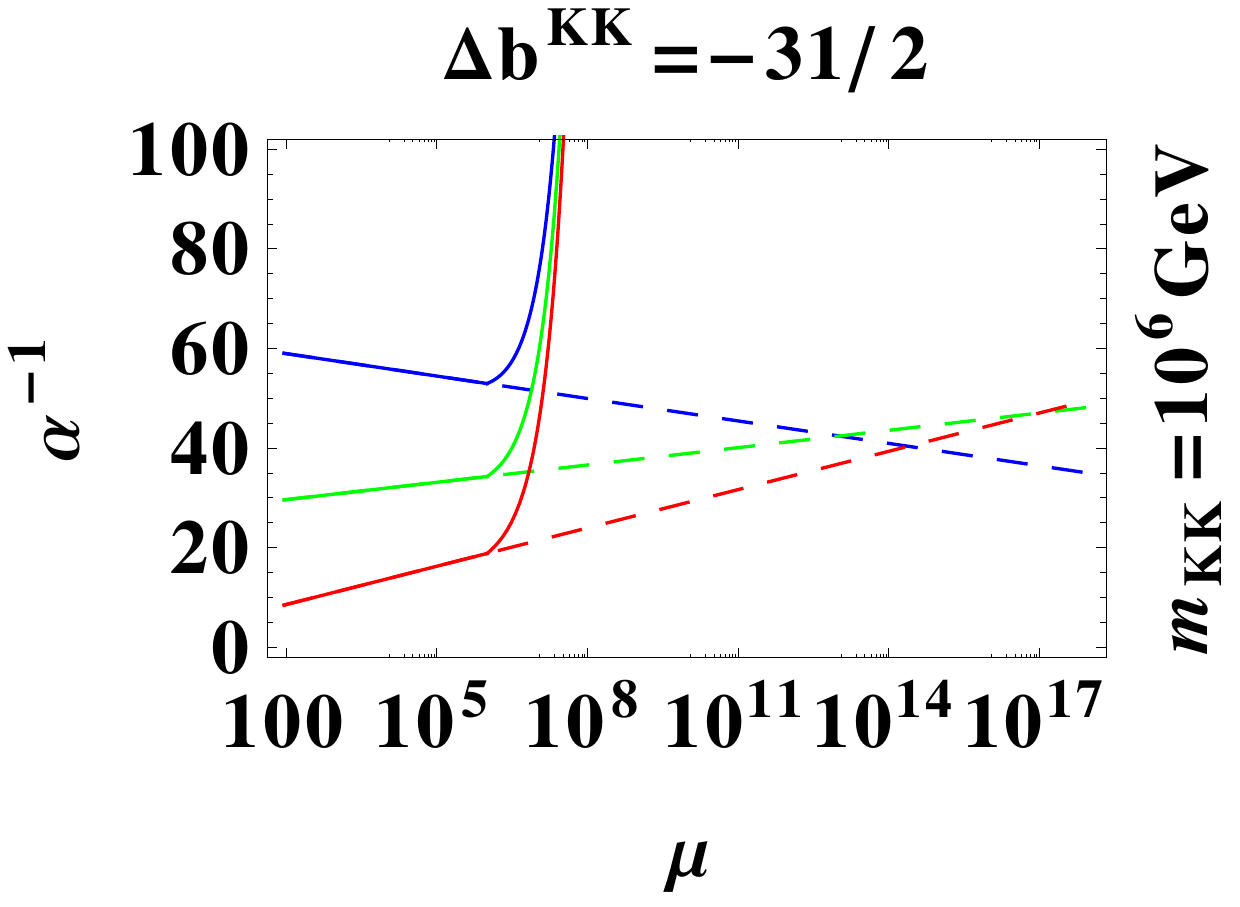}
\includegraphics[bb=0 0 341 248,height=3.5cm]{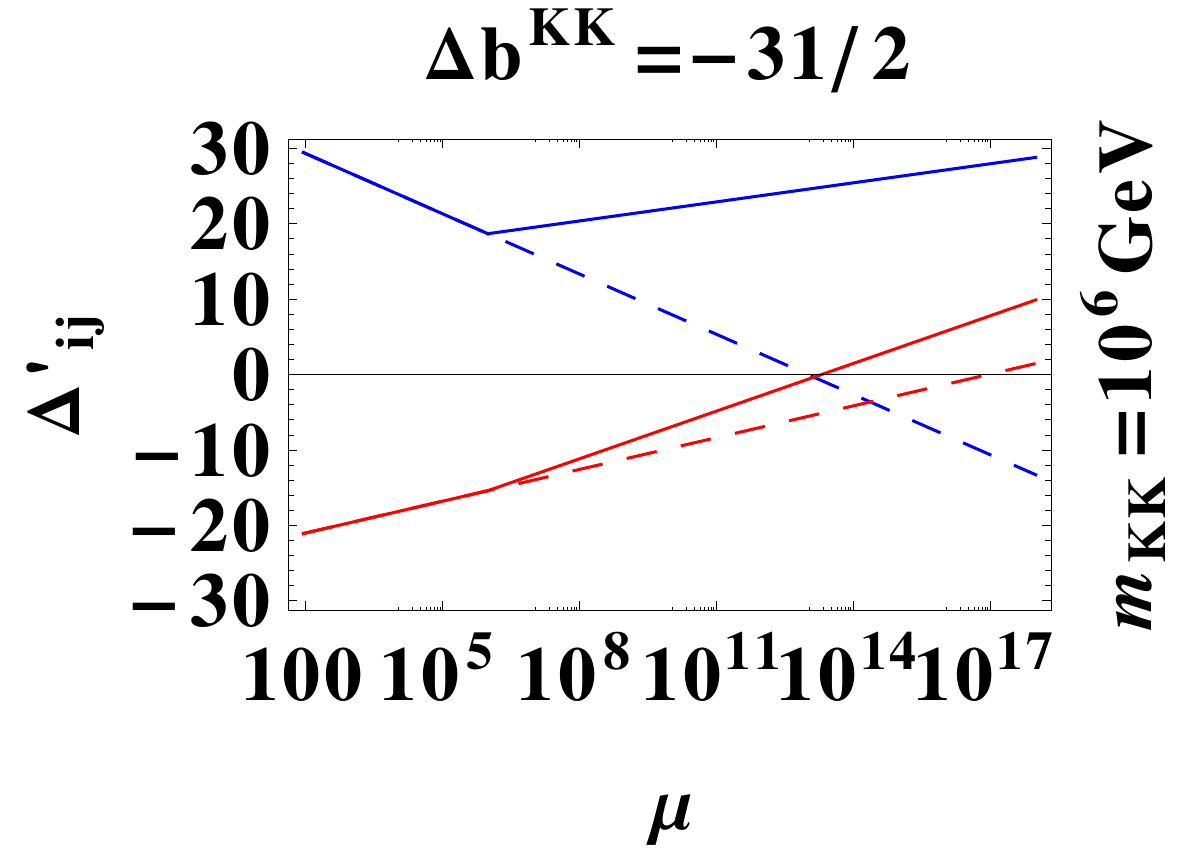}
\includegraphics[bb=0 0 393 247,height=3.5cm]{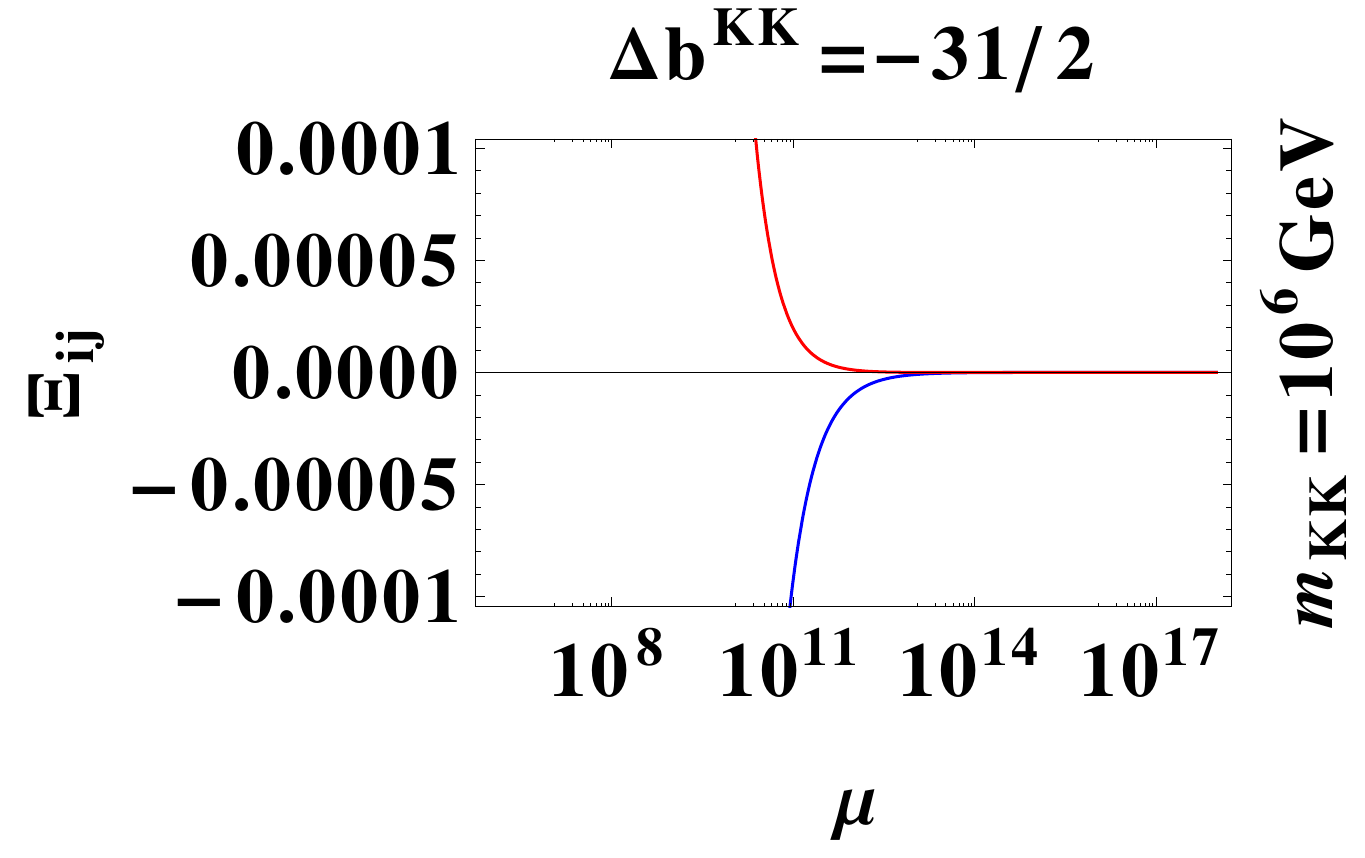}
\end{center}
\caption{$\mu-\alpha^{-1}(\mu)$, $\mu-\Delta'_{ij}(\mu)$,
$\mu-\Xi_{ij}(\mu)$ (Log-Linear plots) in the $SO(11)$ GHGUT with
the matter content 
$(n_{\bf 32},n_{\bf 11})=(3,0)$ ($\Delta b^{KK}=-31/2$)
and KK mass $m_{KK}=10^{6}$ GeV, 
and bulk gauge field mass splitting correction.
For the explanation, see the caption in
Fig.~\ref{Figure:RGE-gauge-coupling-GHGU}.
}
\label{Figure:RGE-gauge-coupling-GHGU-GMS}
\end{figure}

From Fig.~\ref{Figure:RGE-gauge-coupling-GHGU-GMS}, we find the
followings. First, from the center figure $\mu-\Delta'_{ij}(\mu)$ in
Fig.~\ref{Figure:RGE-gauge-coupling-GHGU-GMS}, 
we find that the $\delta b_i^{KK}$ term in
Eq.~(\ref{Eq:4D-gauge-coupling-constant-asymptotic-in-5D-f}) 
is not negligible compared with the $b_i^0$ term, and contributes to 
$\Delta'_{ij}(\mu)$.
From the right figures $\mu-\Xi_{ij}(\mu)$ in
Figs.~\ref{Figure:RGE-gauge-coupling-GHGU} and
\ref{Figure:RGE-gauge-coupling-GHGU-GMS}, the convergence scale is
changed, but this does not affect whether the SM gauge coupling
constants converge or not.
Therefore, we find that 
orbifold BCs or mass spectra affect the detail structure of gauge
couplings described by $\Delta'_{ij}(\mu)$, but
they do not affect the convergence of the SM gauge coupling constants
described by $\Xi_{ij}(\mu)$.

\begin{figure}[tbh]
\begin{center}
\includegraphics[bb=0 0 357 261,height=3.5cm]{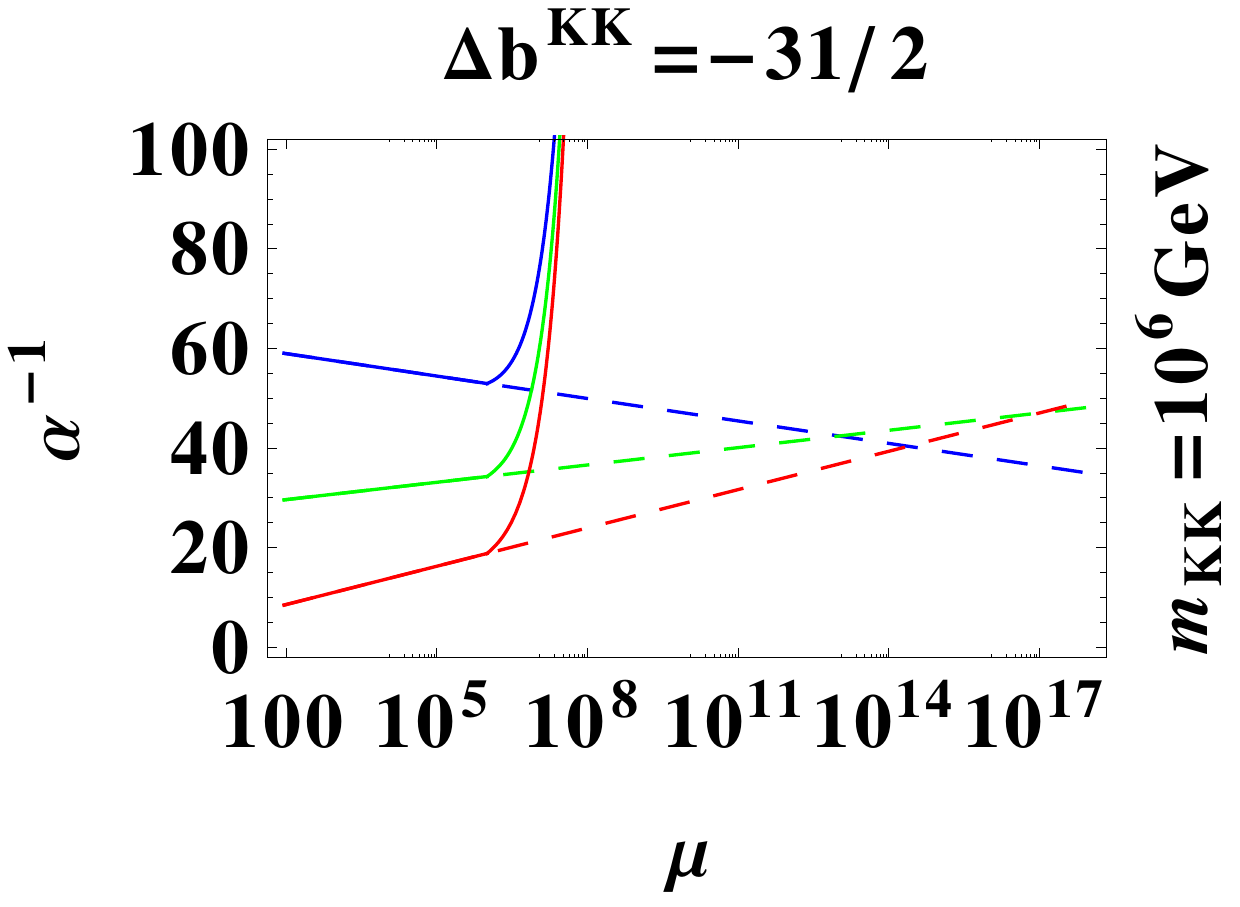}
\includegraphics[bb=0 0 339 247,height=3.5cm]{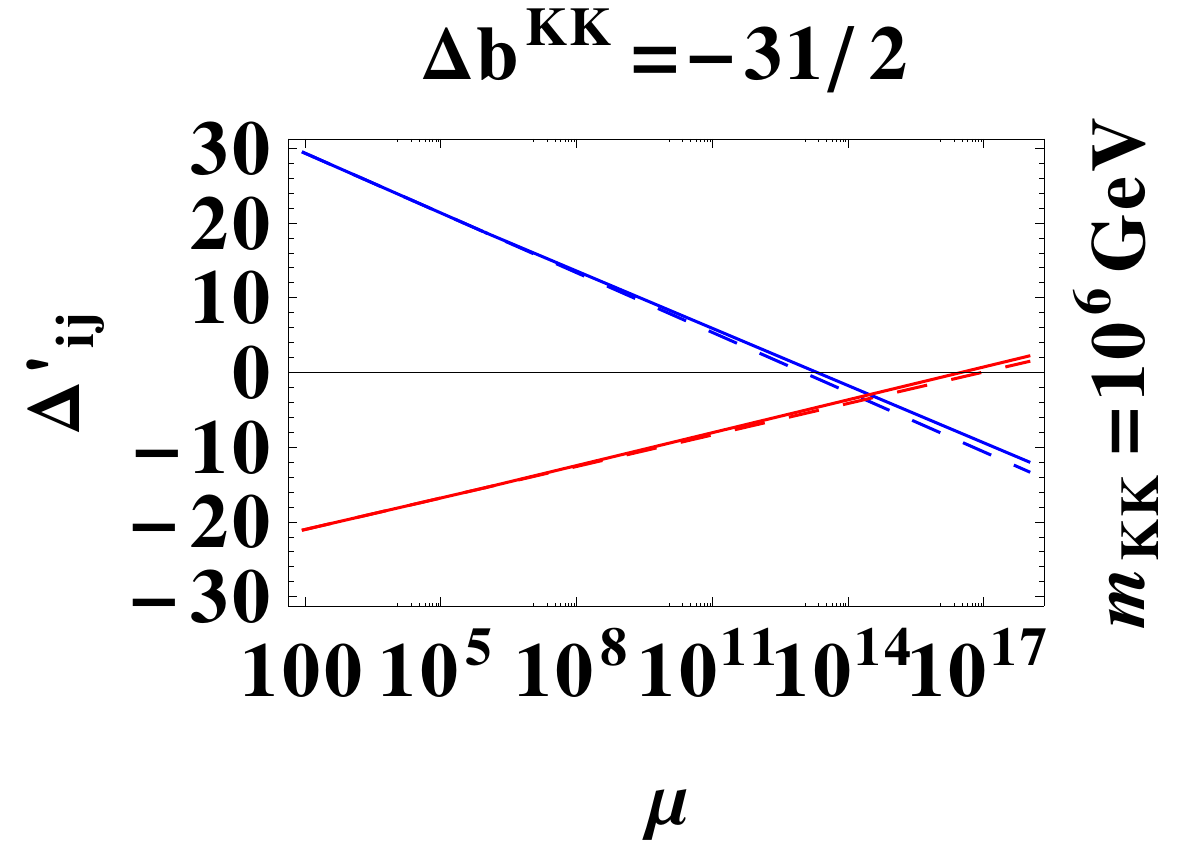}
\includegraphics[bb=0 0 385 242,height=3.5cm]{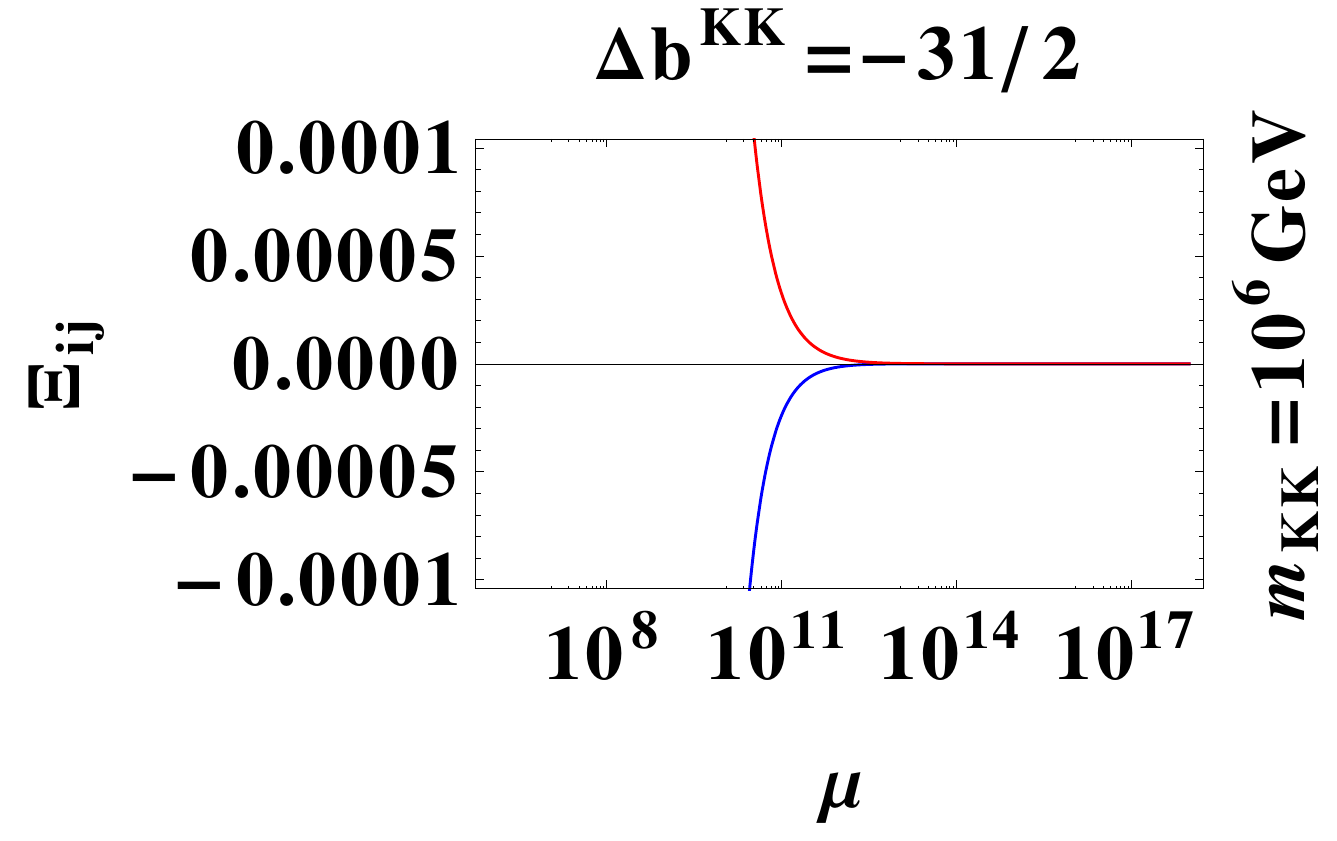}
\end{center}
\caption{$\mu-\alpha^{-1}(\mu)$, $\mu-\Delta'_{ij}(\mu)$,
$\mu-\Xi_{ij}(\mu)$ (Log-Linear plots) in the $SO(11)$ GHGUT with
the matter content 
$(n_{\bf 32},n_{\bf 11})=(3,0)$ ($\Delta b^{KK}=-31/2$)
and KK mass $m_{KK}=10^{6}$ GeV, 
and would-be NG correction.
For the explanation, see the caption in
Fig.~\ref{Figure:RGE-gauge-coupling-GHGU}.
}
\label{Figure:RGE-gauge-coupling-GHGU-wNG}
\end{figure}

We comment on the contribution to RGEs
from the $SO(10)$ spinor brane scalar field on the Planck brane
in Table~\ref{tab:matter-content}.
Its non-vanishing VEV is responsible for
breaking $SO(10)$ to $SU(5)$.  
There are twenty-one would-be NG modes. 
Nine modes are eaten by $G_{PS}/G_{SM}$ gauge bosons, while twelve modes
are uneaten because $SO(10)/G_{PS}$ gauge bosons absorb their
corresponding 5th-dim. components of the 5D gauge
field. The twelve modes become massive via their quantum correction, whose
masses are expected to $O(m_{KK})$ or less depending on dynamics.
They correspond to a complex scalar field with $({\bf 3,2})_{-1/6}$
under $G_{SM}$. It is not any $SU(5)$ multiplet, and it affects the
gauge coupling unification.
The contribution to the $\beta$-function coefficients of $G_{SM}$ is
given by 
\begin{align}
b_i^{\rm wNG}=\frac{1}{3}\sum_{\rm would-be\ NG}T(R_i)
=\left(
\begin{array}{c}
+1/3\\
+1/2\\
+1/5\\
\end{array}
\right),
\end{align}
where this contribution vanishes effectively above the brane mass
scale of $\phi_{\bf 16}$ 
because the $SO(10)$ full multiplet ${\bf 16}$ contribute to 
the $\beta$-function coefficients of $G_{SM}$.
From Fig.~\ref{Figure:RGE-gauge-coupling-GHGU-wNG}, we find that 
it contributes to a gauge coupling unification scale, but the values of
$b_{i}^{\rm wNG}$ are small.

\subsubsection{$m_{KK}<M_{PS}<M_{GUT}\simeq 1/L$}

Let us discuss the Pati-Salam scale $M_{PS}$ effect. In this case, we
have to use different RGEs for the SM and Pati-Salam gauge coupling
constants above and below $M_{PS}$.

\begin{table}[htb]
\begin{center}
\begin{tabular}{|c|cccccc|}\cline{2-7}
\multicolumn{1}{c|}{}&$G_\mu'$&$W_\mu$&$W_\mu'$&$q'$&${u}^{c\prime}$&$\phi'$\\\hline\hline
$SU(4)_C$ &${\bf 15}$&${\bf 1}$&${\bf 1}$&${\bf 4}$&${\bf \overline{4}}$&${\bf 2}$\\ 
$SU(2)_L$  &${\bf 1}$&${\bf 3}$&${\bf 1}$&${\bf 2}$&${\bf 1}$&${\bf 2}$\\ 
$SU(2)_R$  &${\bf 1}$&${\bf 1}$&${\bf 3}$&${\bf 1}$&${\bf 2}$&${\bf 2}$\\ 
$SL(2,\mathbb{C})$&$\left(\frac{1}{2},\frac{1}{2}\right)$&$\left(\frac{1}{2},\frac{1}{2}\right)$&$\left(\frac{1}{2},\frac{1}{2}\right)$&$\left(\frac{1}{2},0\right)$&$\left(\frac{1}{2},0\right)$&$(0,0)$\\[0.5em]
\hline
\end{tabular}
\caption{The SM matter content in the Pati-Salam base.
}
\label{Table:SM-matter-content-in-PS}
\end{center}
\end{table}

We check $\beta$ function coefficients of the Pati-Salam
gauge coupling constants of zero modes by using the RGE in
Eq.~(\ref{Eq:beta-function-coeff-general}).
The matter content of zero mode is shown in
Table~\ref{Table:SM-matter-content-in-PS}.
By using the formula in Eq.~(\ref{Eq:beta-function-coeff-general}) and 
the Dynkin indices listed in
Refs.~\cite{McKay:1981,Slansky:1981yr,Yamatsu:2015gut}, we obtain
\begin{align}
b_i=-\frac{11}{3}C_2(G_i)
+\frac{2}{3}\sum_{\rm Weyl\ Fermions}T(R_i)
+\frac{1}{3}\sum_{\rm Complex\ Scalar}T(R_i)
=\left(
\begin{array}{c}
-32/3\\
-19/6\\
-19/6\\
\end{array}
\right),
\end{align}
where $i=4C,2L,2R$ stand for $SU(4)_C$, $SU(2)_L$, $SU(2)_R$,
respectively.

\begin{table}[htb]
\begin{center}
\begin{tabular}{ccl}
\hline
Field&BC&Representations of $G_{PS}$\\
\hline
$A_\mu$&
 $(N,N)$&$({\bf 15,1,1})$, $({\bf 1,3,1})$, $({\bf 1,1,3})$\\
&$(N,D)$&$({\bf 6,2,2})$\\
&$(D,N)$&$({\bf 6,1,1})$\\
&$(D,D)$&$({\bf 1,2,1})$\\
$A_y$&
 $(N,N)$&$({\bf 1,2,2})$\\
&$(N,D)$&$({\bf 6,1,1})$\\
&$(D,N)$&$({\bf 6,2,2})$\\
&$(D,D)$&$({\bf 15,1,1})$, $({\bf 1,3,1})$, $({\bf 1,1,3})$\\
\hline
\end{tabular}
\caption{The orbifold BCs of the components of the $SO(11)$
 bulk gauge field $A_M=A_\mu\oplus A_y$ in the Pati-Salam base.
}
\label{Table:SO(11)-bulk-gauge-field-BCs-PS}
\end{center}
\end{table}

We consider the contribution to $\delta b_{i}^{KK}$ from the mass
spectra of the $SO(11)$ bulk gauge field. 
As we discussed before, the would-be NG bosons do not affect the RGEs
for the SM gauge coupling constants.
We can calculate $\delta b_{i}^{KK}$ $(i=4C,2L,2R)$ by using 
the orbifold BCs of the $SO(11)$ bulk gauge field shown in 
Table~\ref{Table:SO(11)-bulk-gauge-field-BCs-PS}:
\begin{align}
\delta b_{4C}^{KK}=-\frac{35}{3},\ \
\delta b_{2L}^{KK}=-21,\ \
\delta b_{2R}^{KK}=-21.
\end{align}

We have to use the RGEs for three SM gauge coupling constants
below $M_{PS}$, while we have to use the RGEs for three Pati-Salam gauge
coupling constants. To connect them, we use the following matching
condition at the Pati-Salam scale $M_{PS}$ $(m_{KK}<M_{PS}<M_{GUT})$,
\begin{align}
\alpha_{3C}(M_{PS})&=\alpha_{4C}(M_{PS}),\\
\alpha_{2L}(M_{PS})&=\alpha^\prime_{2L}(M_{PS}),\\
\alpha^{-1}_{1Y}(M_{PS})&=
\frac{3}{5}\alpha^{-1}_{2R}(M_{PS})+\frac{2}{5}\alpha^{-1}_{4C}(M_{PS}),
\end{align}
where they are determined by the normalization conditions of the
generators of $G_{PS}$ and $G_{SM}$.
(See e.g., Ref.~\cite{Mohapatra2002} in detail.)

\begin{figure}[tb]
\begin{center}
\includegraphics[bb=0 0 341 248,height=3.5cm]{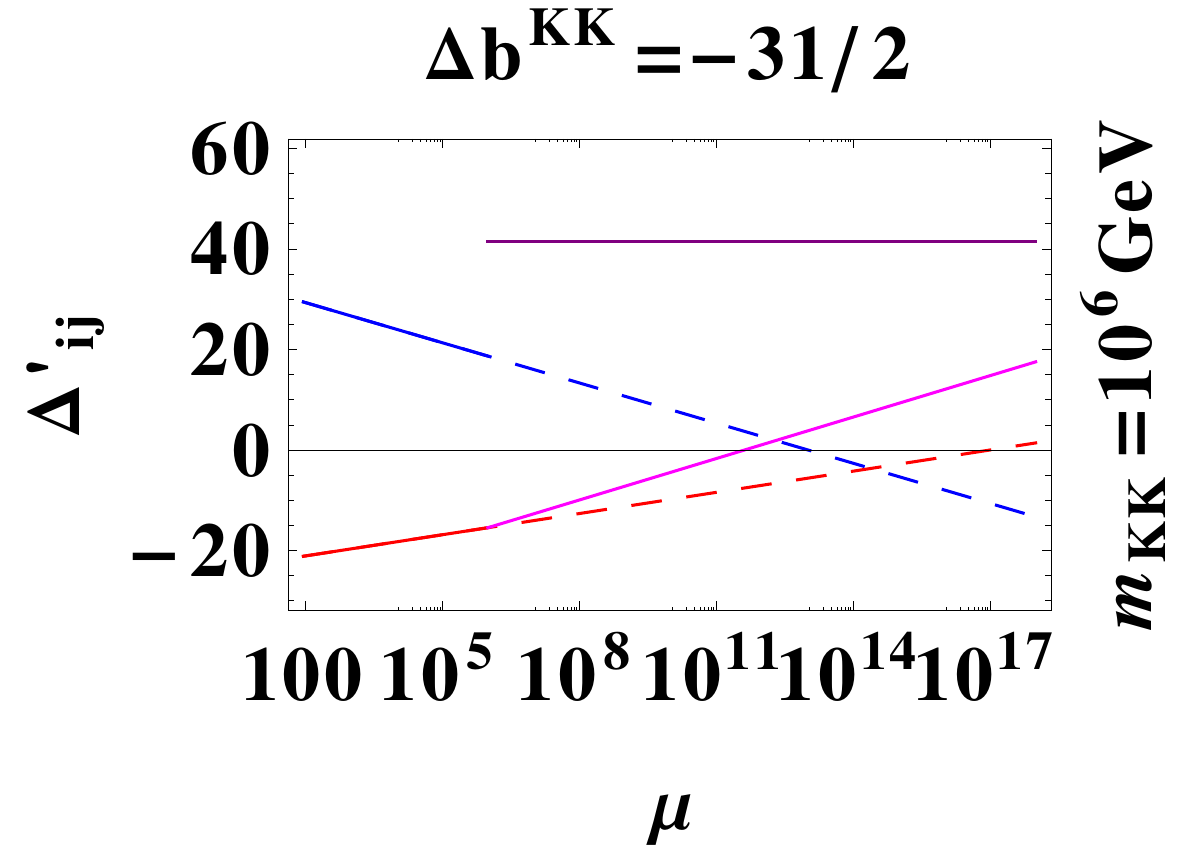}
\includegraphics[bb=0 0 339 247,height=3.5cm]{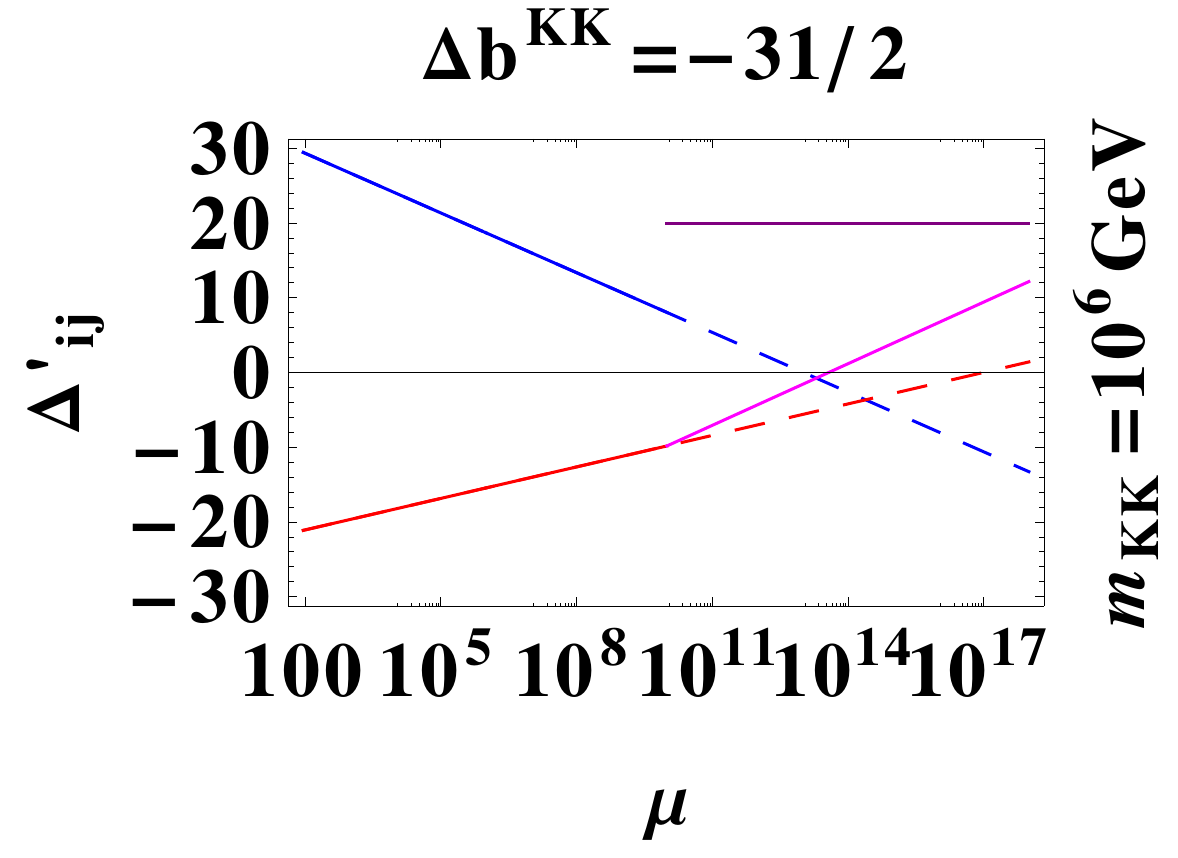}
\includegraphics[bb=0 0 335 244,height=3.5cm]{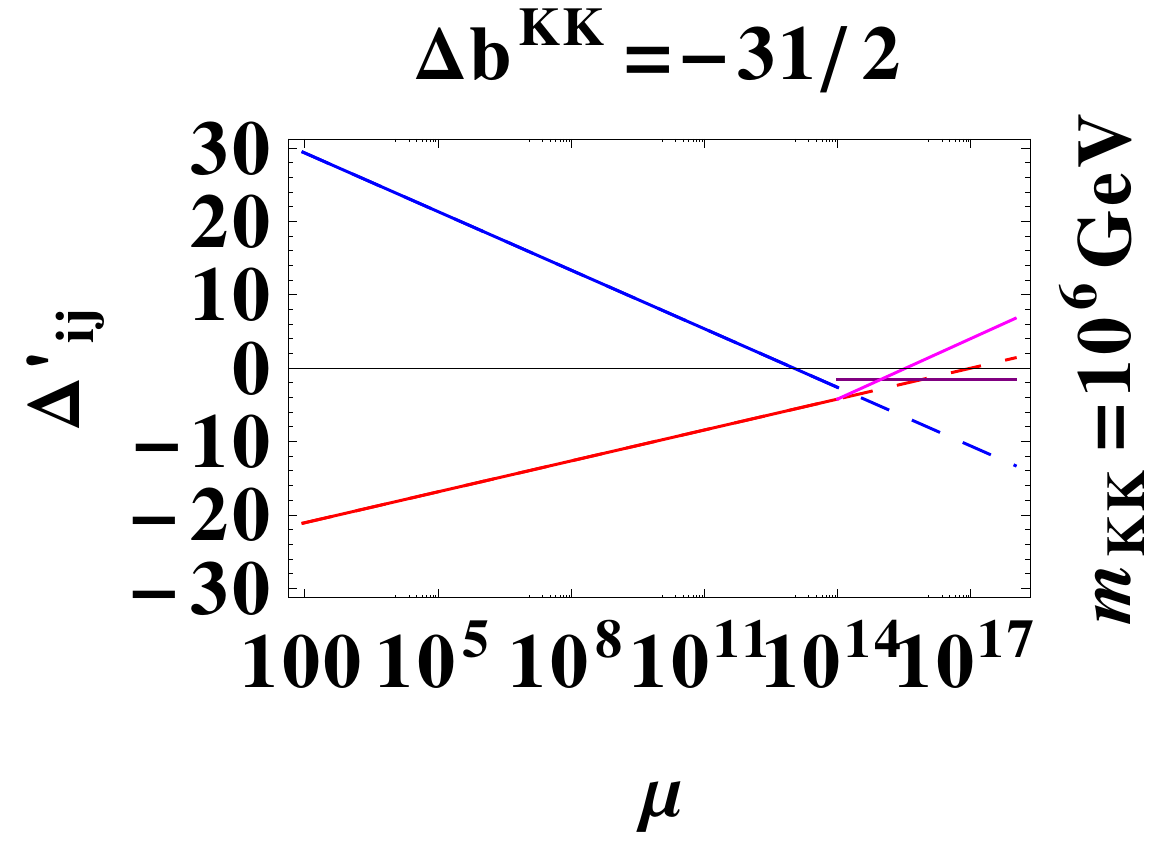}\\
\includegraphics[bb=0 0 353 255,height=3.5cm]{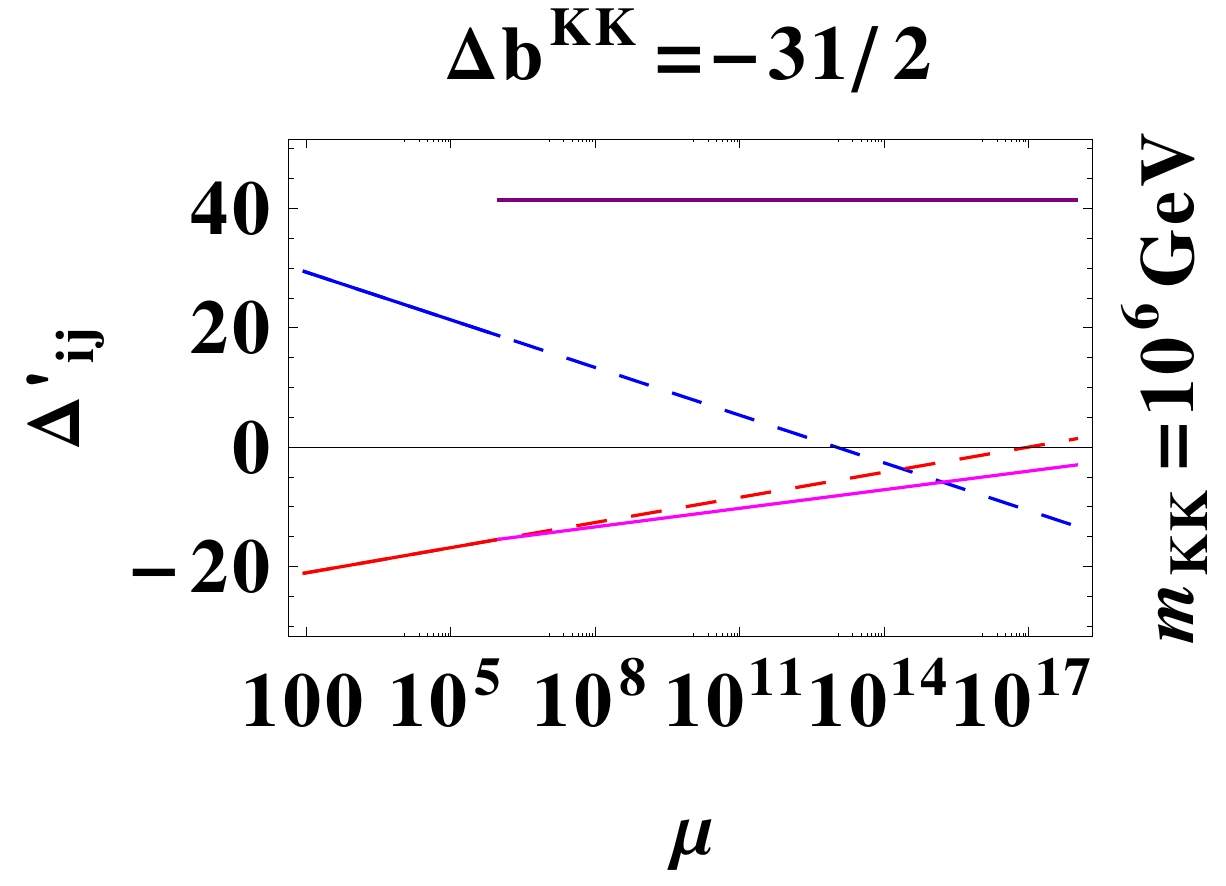}
\includegraphics[bb=0 0 341 248,height=3.5cm]{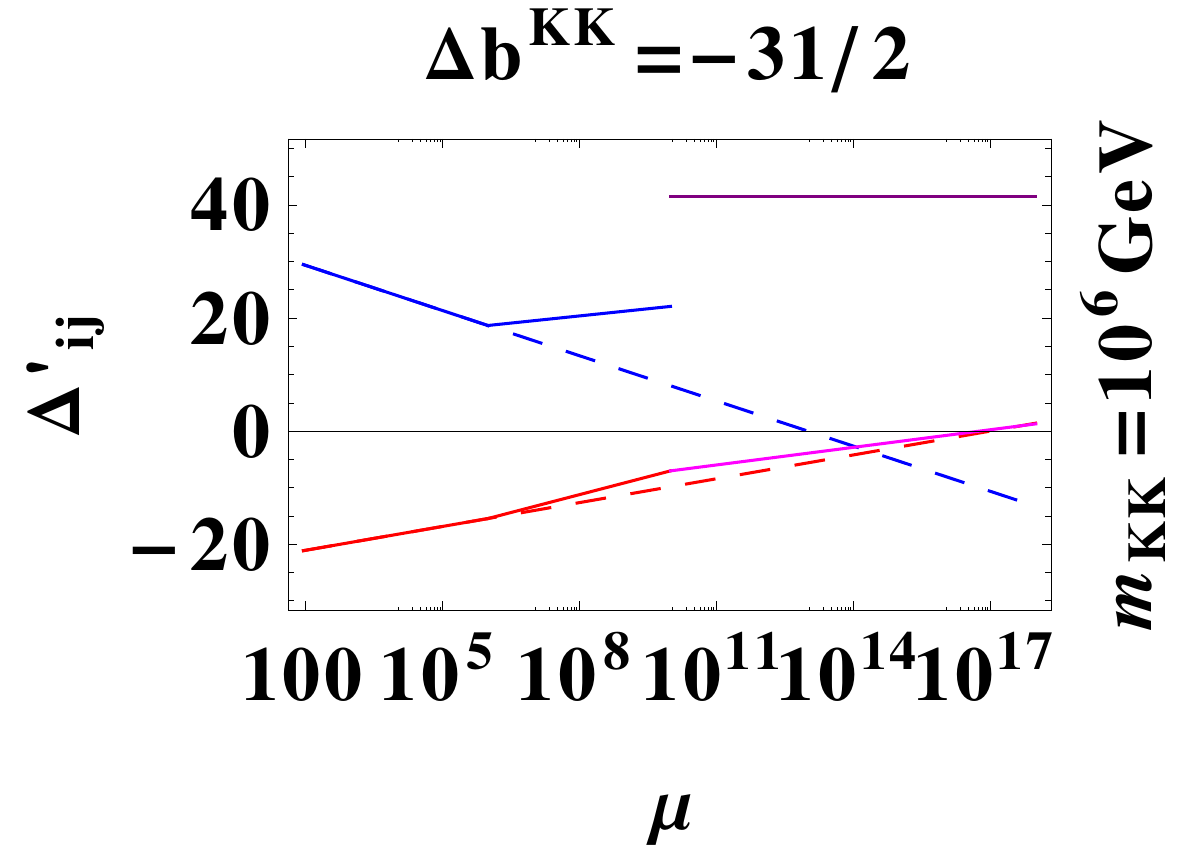}
\includegraphics[bb=0 0 336 245,height=3.5cm]{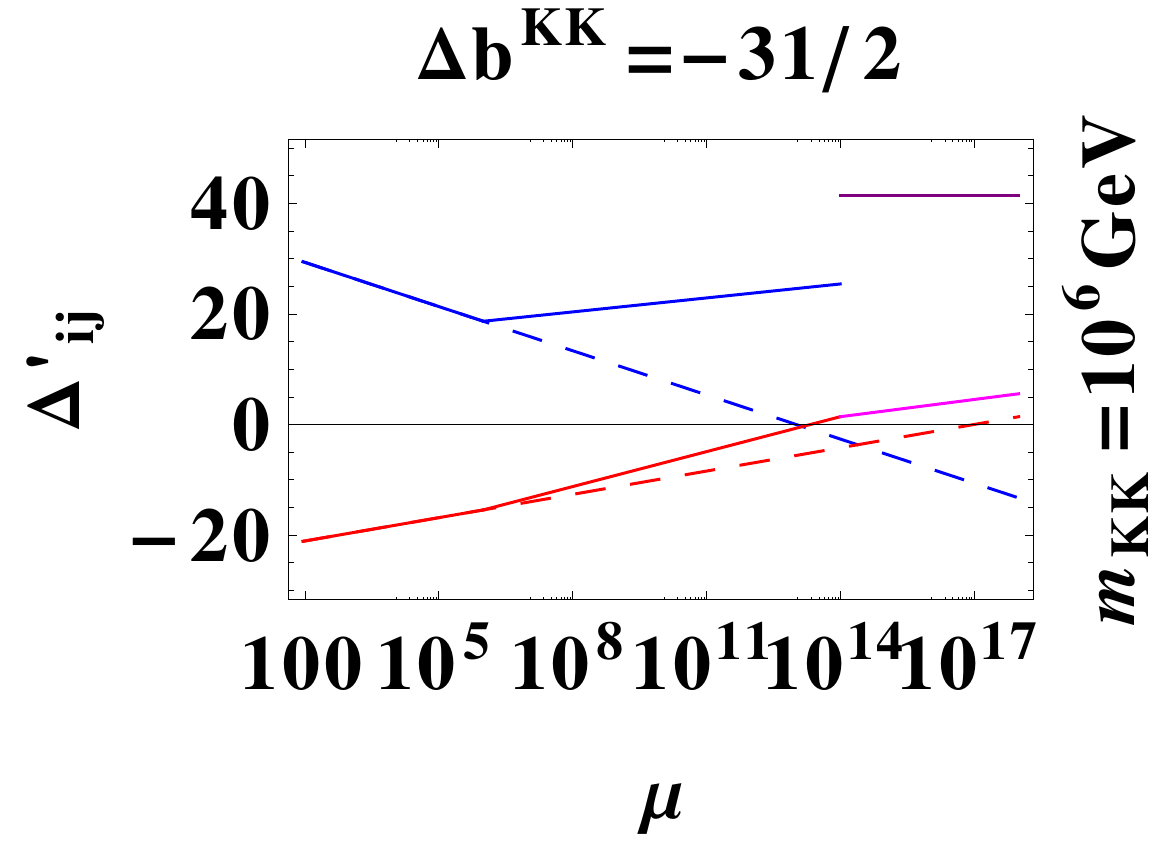}
\end{center}
\caption{$\mu-\Delta'_{ij}(\mu)$ (Log-Linear plots) in $SO(11)$ GHGUT
with the same matter content 
$(n_{\bf 32},n_{\bf 11})=(3,0)$ ($\Delta b^{KK}=-31/2$)
and KK mass $m_{KK}=10^{6}$ GeV, 
and Pati-Salam scales $M_{PS}=10^{6},10^{10},10^{14}$ GeV: 
the top figures do not include the $SO(11)$ bulk gauge field mass
splitting correction;
the bottom figures include the $SO(11)$ bulk gauge field mass splitting
correction. 
The red line is $\Delta'_{3C,2L}=\alpha_{3C}^{-1}-\alpha_{2L}^{-1}$,
the blue line is $\Delta'_{1Y,2L}=\alpha_{1Y}^{-1}-\alpha_{2L}^{-1}$,
the purple line is $\Delta'_{2R,2L}=\alpha_{2R}^{-1}-\alpha_{2L}^{-1}$,
and 
the magenta line is $\Delta'_{4C,2L}=\alpha_{4C}^{-1}-\alpha_{2L}^{-1}$,
}
\label{Figure:RGE-gauge-coupling-GHGU-PS-Delta}
\end{figure}

From the $\mu-\Delta'_{ij}(\mu)$ figures in
Fig.~\ref{Figure:RGE-gauge-coupling-GHGU-PS-Delta},
we find that the Pati-Salam scale w/o the orbifold BCs (mass splitting) 
affect the detail structure of gauge couplings described by
$\Delta'_{ij}(\mu)$. Thus, even when we take into account 
the Pati-Salam scale, the orbifold BCs, etc., they do not change our
discussion about asymptotic freedom of the SM gauge coupling constants
and gauge coupling unification.
For $m_{KK}=10^{6}$ GeV and $\mu=10^{11-12}$, 
$\mbox{Err}[\Delta'_{ij}(10^{11-12}\mbox{GeV})]\simeq O(10-100)$ and
the deviations $\Delta'_{3C,2L}(10^{11-12}\mbox{GeV})$ 
and $\Delta'_{1Y,2L}(10^{11-12}\mbox{GeV})$ are less than 50 from the
center figure in Fig.~\ref{Figure:RGE-gauge-coupling-GHGU-PS-Delta}, and
then $M_{GCU}$ starts around $10^{11-12}$ GeV.

\section{Summary and discussion}
\label{Sec:Summary-discussion}

We discussed the RGEs for the 4D SM gauge coupling constants in the
$SO(11)$ gauge-Higgs grand unification scenario on the 5D RS warped
spacetime. We found that the 4D SM gauge coupling constants are
asymptotically free in the $SO(11)$ GHGUTs with the matter contents shown in
Tables~\ref{Table:three-chiral-generations-asymptotic-free-SO11} and 
\ref{Table:three-chiral-generations-asymptotic-free-fermion-number-SO11},
which satisfy $\Delta b^{KK}<0$.
We also discussed the SM gauge coupling unification.
We showed that the three SM gauge coupling constants are effectively
unified above the almost SM gauge coupling unification scale $M_{GCU}$
discussed in Sec.~\ref{Sec:Asymptotic-freedom-Gauge-coupling-unification}.
We have not fixed the GUT or compactification scale $M_{GUT}=1/L$,
but as long as $M_{GUT}=1/L$ is larger than $M_{GCU}$, there is no any
inconsistency within at least the current experimental accuracy of the
SM gauge coupling constants.
In Sec.~\ref{Sec:Corrections}. we showed that the correction from the
mass spectra of the $SO(11)$ bulk gauge fields, the would-be NG boson,
and the Pati-Salam scale does not affect the asymptotic freedom and
gauge coupling unification of the SM gauge couplings, while they affect
the detail structures of the RGE running.
From the above, we find that the Weinberg angle at $\mu=M_{GUT}$,
$\sin^2\theta_W(M_{GUT})=3/8$, is consistent with that at $\mu=M_{Z}$,
$\sin^2\theta_W(M_{Z})\simeq 0.23$.

In this paper, we mainly considered the $SO(11)$ GHGUTs, but our
discussion can be applied for other GHGUTs.
E.g., we have already found the asymptotic freedom condition for a gauge
coupling constant in general GHGUTs based on any simple Lie group $G$
in Eq.~(\ref{Eq:Condition-asymptotic-free}). It is very easy to list up
the the matter contents that satisfy the asymptotic freedom condition by
using Tables in Ref.~\cite{Yamatsu:2015gut}.

We discussed the RGEs for the 4D SM gauge coupling constants in 5D RS
warped spacetime by using the KK expansion. There is another approach
about them by using AdS/CFT-like correspondence
in Refs.~\cite{Randall:2001gc,Randall:2001gb,Goldberger:2002cz,Goldberger:2002hb,Goldberger:2002pc}. 

\section*{Acknowledgments}

The author would like to thank Yutaka Hosotani for many stimulus
discussions and valuable comments, and also thank Hidenori Fukaya,
Taichiro Kugo, Minoru Tanaka, and Satoshi Yamaguchi for useful comments. 
This work was supported in part by Japan Society for the Promotion of
Science Grants-in-Aid for Scientific Research No.~23104009.

\bibliographystyle{utphys} 
\bibliography{../../arxiv/reference}

\end{document}